\documentclass[prd,aps,floats,superscriptaddress,preprintnumbers,preprint]{revtex4}

\usepackage[latin1]{inputenc}                    
\usepackage{graphicx}                            
\usepackage{latexsym}                            
\usepackage{amsfonts}                            
\usepackage{amssymb}                             
\usepackage{amsmath}                             
\usepackage[mathscr]{eucal}                      
\usepackage{dcolumn}                             
\usepackage{theorem}                             
\usepackage{footnote}
\usepackage{enumitem}
\usepackage{bm}
\usepackage{color}

\def\be{\begin{equation}}
\def\ee{\end{equation}}
\def\bea{\begin{eqnarray}}
\def\eea{\end{eqnarray}}
\def\bf{\begin{figure}}
\def\ef{\end{figure}}
\def\bc{\begin{center}}
\def\ec{\end{center}}

\def\gg{\gamma \gamma}
\def\gZ{\gamma Z}
\def\qw{$Q_W^p \, $}

\def\qwe{$Q_{\text{weak}} \,\,$}
\def\sw{\sin^2\theta_W}
\def\bgZ{$\square_{\gZ}\, $}
\def\bgZv{$\square_{\gZ}^V \,$}
\def\bgZa{$\square_{\gZ}^A \,$}
\def\regzv{$\Re e \, \square_{\gZ}^V\ $}
\def\imgzv{$\Im m \, \square_{\gZ}^V\ $}

\def\stl{\sigma_{T,L}}
\def\nn{\nonumber}
\def\ftgg{F_2^{\gg}}
\def\fogg{F_1^{\gg}}
\def\flgg{F_L^{\gg}}
\def\figg{F_i^{\gg}}
\def\ftgz{F_2^{\gZ}}
\def\fogz{F_1^{\gZ}}
\def\flgz{F_L^{\gZ}}
\def\fthgz{F_3^{\gZ}}
\def\figz{F_i^{\gZ}}

\def\stl{\sigma_{T,L}}

\begin{document}

\preprint{
\vbox{
\hbox{ADP-12-19/T786, JLAB-THY-13-1698}
}}

\title{Constrained $\gZ$ interference corrections
	to parity-violating electron scattering}
\author{N.~L.~Hall}
\affiliation{\mbox{ARC Centre of Excellence for Particle Physics 
	at the Terascale, and CSSM}, School of Chemistry and Physics, 
	University of Adelaide, Adelaide, South Australia 5005, Australia}
\author{P.~G.~Blunden}
\affiliation{\mbox{Department of Physics and Astronomy,
	University of Manitoba}, Winnipeg,  Manitoba, Canada R3T 2N2}
\author{W.~Melnitchouk}
\affiliation{\mbox{Jefferson Lab, 
	12000 Jefferson Avenue, Newport News, Virginia 23606, USA}}
\author{A.~W.~Thomas}
\affiliation{\mbox{ARC Centre of Excellence for Particle Physics 
	at the Terascale, and CSSM}, School of Chemistry and Physics, 
	University of Adelaide, Adelaide SA 5005, Australia}
\author{R.~D.~Young}
\affiliation{\mbox{ARC Centre of Excellence for Particle Physics 
	at the Terascale, and CSSM}, School of Chemistry and Physics, 
	University of Adelaide, Adelaide SA 5005, Australia}

\date{\today}

\begin{abstract}
We present a comprehensive analysis of $\gZ$ interference corrections
to the weak charge of the proton measured in parity-violating electron
scattering, including a survey of existing models and a critical
analysis of their uncertainties.  Constraints from parton distributions
in the deep-inelastic region, together with new data on parity-violating
electron scattering in the resonance region, result in significantly
smaller uncertainties on the corrections compared to previous estimates.
At the kinematics of the \qwe experiment, we determine the $\gZ$ box
correction to be
	$\Re e\, \square_{\gZ}^V = (5.57 \pm 0.36) \times 10^{-3}$.
The new constraints also allow precise predictions to be made for
parity-violating deep-inelastic asymmetries on the deuteron.
\end{abstract}



\maketitle

%
\section{Introduction}

Modern low-energy experiments at the precision frontier provide
important alternatives to high-energy tests of the Standard Model
currently being performed at the Large Hadron Collider (for recent
reviews, see Refs.~\cite{Sirlin:2012mh, Kumar:2013yoa, Erler:2013xha}).
One such experiment is the parity-violating (PV) elastic
electron-proton scattering measurement that was recently carried
out by the \qwe collaboration at Jefferson Lab \cite{Armstrong:2012ps},
which aims to determine the proton's weak charge \qw\! to within 4\%. 
At tree level, the weak charge is related to the weak mixing angle,
$\sw$, by $Q_W^p = 1 - 4 \sw$.
By scattering low-energy polarized electrons from an unpolarized
hydrogen target, \qwe measured the asymmetry between the cross
sections for right- and left-handed electrons,
\bea
A_{\rm PV} &=& \frac{\sigma_+ - \sigma_-}{\sigma_+ + \sigma_-},
\eea
where $\sigma_\lambda$ is the cross section for a right-hand (helicity
$\lambda = +1$) or left-hand (helicity $\lambda = -1$) electron.
At small four-momentum transfer squared $t$, the asymmetry is related
to $Q_W^p$ by \cite{Musolf:1993tb}
\bea
A_{\rm PV} &=& \frac{G_F}{4 \pi \alpha \sqrt{2}}\, t\, Q_W^p,
\eea
where $G_F$ is the Fermi constant and $\alpha$ is the fine structure
constant.
Including radiative corrections, the proton's weak charge can be
written as \cite{Erler:2003yk}
\bea
Q_W^p &=& \left( 1 +\Delta\rho + \Delta_e \right)
	  \left( 1 - 4 \sin^2\theta_W(0) + \Delta_e^{'} \right)
       + \square_{WW} + \square_{ZZ} + \square_{\gZ}(0), 
\label{eq:qwHO}
\eea
where $\sin^2\theta_W(0)$ is the weak mixing angle at zero momentum,
and the correction terms $\Delta \rho$, $\Delta_e$ and $\Delta_e^{'}$
are well understood and have been computed to sufficient levels of
precision \cite{Erler:2003yk}.  Similarly, the work of
Refs.~\cite{Marciano:1982mm, Marciano:1983ss, Musolf:1990ts} has
established that the electroweak box diagrams $\square_{WW}$ and
$\square_{ZZ}$ are known within \qwe uncertainty limits.

\begin{figure}[tb]
\includegraphics[width=6.5cm]{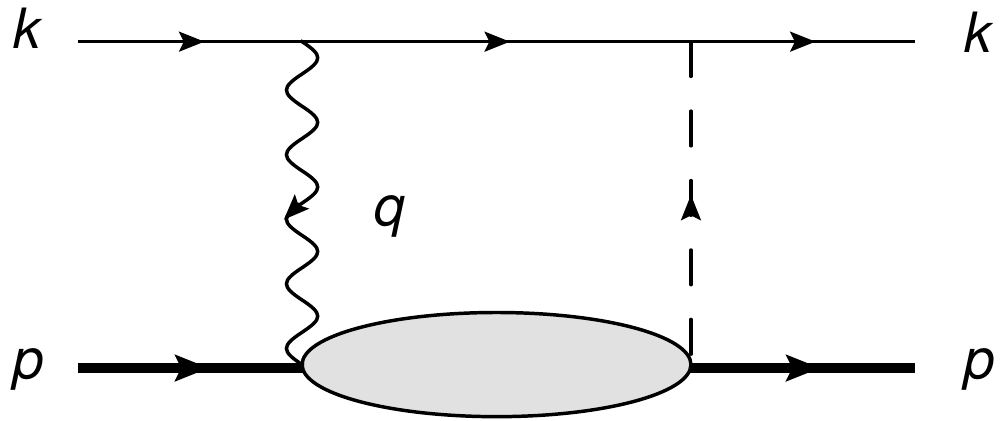}\hspace*{1.7cm}
\includegraphics[width=6.5cm]{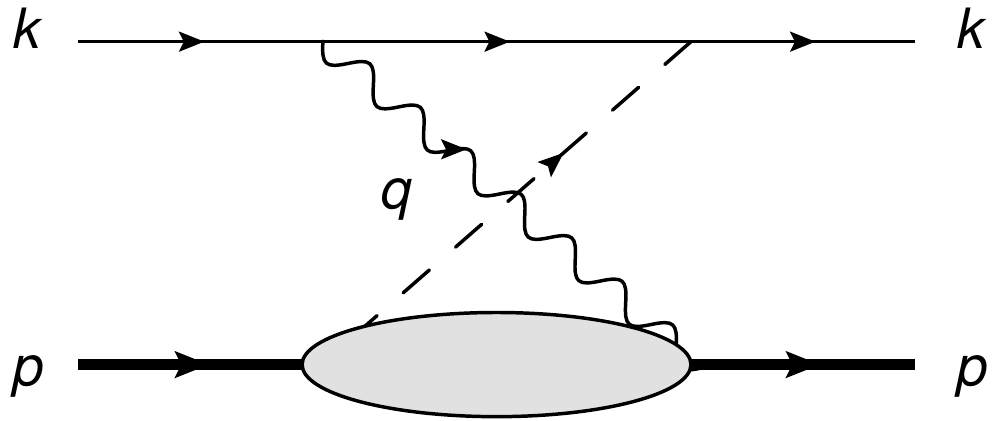}
\caption{Interference $\gZ$ box (left) and crossed box (right)
	diagrams.  The wavy and dashed lines represent the exchanged
	$\gamma$ and $Z$ bosons, with the electron, hadron and virtual
	photon momenta labeled by $k$, $p$, and $q$, respectively.}
\label{fig:gZbox}
\end{figure}

Until recently it was also believed that the interference $\gZ$
contribution, illustrated in Fig.~\ref{fig:gZbox}, was known to
sufficient accuracy for the \qwe experiment.
This correction is defined in terms of the electroweak amplitudes as
\cite{Arrington:2011dn}
\bea
\square_{\gZ}(0)
&=& Q_W^p\,
 { \Re e \left( {\cal M}_\gamma^* {\cal M}_{\gZ}^{\rm (PV)} \right)
   \over
   \Re e \left( {\cal M}_\gamma^* {\cal M}_Z^{\rm (PV)} \right) },
\label{eq:gZbox}
\eea
where ${\cal M}_\gamma$ is the electromagnetic Born amplitude,
${\cal M}_{Z}^{\rm (PV)}$ is the parity-violating part of the
Born $Z$ exchange amplitude, and ${\cal M}_{\gZ}^{\rm (PV)}$
is the parity-violating part of the $\gZ$ interference amplitude
(including the contributions with the $\gamma$ and $Z$ interchanged).
A groundbreaking contribution was made by Gorchtein and Horowitz
\cite{Gorchtein:2008px}, who showed, using a dispersion relations
approach, that the \bgZ term was strongly energy dependent and was
much larger at \qwe energies ($\sim 1$~GeV) than previous estimates
had assumed \cite{Erler:2003yk}.  More importantly, the uncertainty
on this correction was such that it could significantly affect the
precision aims of the \qwe measurement.

Subsequent analyses by Sibirtsev {\it et al.} \cite{Sibirtsev:2010zg}
and Rislow and Carlson \cite{Rislow:2010vi} generally agreed with the
overall scale of the correction found in Ref.~\cite{Gorchtein:2008px},
but disputed the magnitude of the uncertainties.  In a follow-up
study, Gorchtein {\it et al.} \cite{Gorchtein:2011mz} performed
a more detailed analysis of the model dependence of the \bgZ
contribution, correcting several errors from the original analysis
\cite{Gorchtein:2008px}, but still quoted uncertainties twice as
large as those in Refs.~\cite{Sibirtsev:2010zg, Rislow:2010vi}.

Since the interpretation of the \qwe results depends on having a sound
understanding of the \bgZ correction, the lack of consensus about the
magnitude of its uncertainty is obviously problematic.  To move beyond
this impasse, in this paper we revisit this problem with the aim of
resolving the disagreements.

We begin our discussion by outlining in Sec.~\ref{sec:disp} the
dispersion relation formalism used to compute the $\gZ$
corrections in terms of $\gZ$ interference structure functions.
The latter are the main input into the calculations and are reviewed
in detail in Sec.~\ref{sec:SFs}.  In particular, we discuss the
uncertainties in determining the $\gZ$ structure functions
from electromagnetic data for both the resonance and nonresonant
background contributions.
Constraints from parton distribution functions in the deep-inelastic
scattering (DIS) region and new data from the parity-violating
electron-deuteron scattering experiment E08-011 at Jefferson Lab 
\cite{E08011-RES} in the resonance region are used in
Sec.~\ref{sec:constraints} to limit the uncertainty range in models
for the $\gZ$ structure functions, and to provide more reliable
bounds on the box corrections.
The resulting $\square_{\gZ}^V$ correction is presented in
Sec.~\ref{sec:res}, where we contrast the revised uncertainties
with those estimated in previous unconstrained analyses.
Predictions are also made for parity-violating deuteron asymmetries
in the deep-inelastic region, as well as for the recently completed
inelastic measurement by the \qwe Collaboration \cite{Qweak-inel}.
Finally, we draw some general conclusions from this analysis in
Sec.~\ref{sec:conc} and explore possibilities to further reduce
the uncertainties on the $\gZ$ corrections in the future.

\section{Dispersive analysis of parity-violating electron-hadron scattering}
\label{sec:disp}

The $\gZ$ interference correction $\square_{\gZ}$ can be
decomposed into two parts, arising from the electron vector with
hadronic axial-vector coupling to the $Z$ boson ($\square_{\gZ}^A$)
and from the electron axial-vector with vector hadronic coupling 
to the $Z$ ($\square_{\gZ}^V$):
\bea
\square_{\gZ}(E) &=& \square_{\gZ}^A(E)\ +\ \square_{\gZ}^V(E).
\label{eq:gZ}
\eea
At very low energies, such as those relevant for atomic parity violation
experiments \cite{Wood:1997zq, Dzuba:2012bh}, the \bgZa term dominates,
while the contribution from the \bgZv is negligible.  At the energy of
the \qwe experiment, however, both terms provide significant
contributions.
The \bgZa corrections were first computed some time ago by Marciano
and Sirlin \cite{Marciano:1982mm, Marciano:1983ss} and were updated
recently within a dispersion relation framework by Blunden {\it et al.}
\cite{Blunden:2011rd, Blunden:2012ty}, with reduced errors.
The vector hadron correction, \bgZv, which is subject to significantly
larger uncertainty, will be the focus of the rest of this analysis.
We will consider only the inelastic contribution to \bgZ;
the elastic contribution has previously been considered in
Refs.~\cite{Marciano:1982mm, Marciano:1983ss, Zhou:2007hr, Tjon:2009hf}
and is strongly suppressed by an additional factor $Q_W^p$.

For forward scattering, the dispersion relation for the real part of
\bgZv is given by
\bea
\Re e\, \square_{\gZ}^V (E)
%
%
&=& \frac{2E}{\pi}
  {\cal P} \int_0^\infty dE' {1 \over E'^2-E^2}\,
  \Im m\, \square_{\gZ}^V(E'),
\label{eq:DR}
\eea
where ${\cal P}$ denotes the principal value integral, and we
have used the fact that \bgZv is odd under the interchange
$E' \leftrightarrow -E'$.  From the optical theorem, the imaginary
part of the PV $\gZ$ exchange amplitude can be written as
\cite{Gorchtein:2008px, Sibirtsev:2010zg, Arrington:2011dn}
\bea
2\, \Im m\, {\cal M}_{\gZ}^{\rm (PV)}
&=& -4 \sqrt{2} \pi M G_F
    \int\!\!{d^3k' \over (2\pi)^3 2 E_{k'}}
    \left( { 4 \pi \alpha \over Q^2} \right)
    {1 \over 1+Q^2/M_Z^2}\,
    L_{\mu\nu}^{\gZ}\, W^{\mu\nu}_{\gZ},
\label{eq:ImBoxdef}
\eea
where $Q^2 = -q^2$ represents the virtuality of the exchanged
boson, and the integration variable $k' = k - q$.
The $\gZ$ lepton tensor is given by
\bea
L_{\mu\nu}^{\gZ}
&=& \bar{u}(k,\lambda)\,
    (g_V^e \gamma_\mu - g_A^e \gamma_\mu \gamma_5)\!
    \not\!k'\, \gamma_\nu\, u(k,\lambda),
%
%
\label{eq:Lep}
\eea
where the vector and axial-vector couplings of the electron to
the weak current are $g^e_V = -(1 - 4 \sw)/2$ and $g^e_A = -1/2$,
respectively, and $\lambda$ is the lepton helicity.
The hadronic tensor for a nucleon initial state is defined as
\bea
\hspace*{-0.3cm}
W^{\mu\nu}(p,q)\
&=& {1 \over 2 M} \sum_X 
  \langle N (p)  | J^\mu(0) | X (p_X) \rangle
  \langle X (p_X) | J^\nu(0) | N (p) \rangle
  (2 \pi)^3 \delta^{(4)}(q + p - p_X),
\eea
where $J_\gamma^\mu$ and $J_Z^\mu$ are the electromagnetic
and weak neutral currents, respectively, and $p_X$ is the
four-momentum of the hadronic intermediate state $X$.
Using isospin symmetry, the matrix elements of the vector
component of the $Z$ current for a proton target can be
related to the proton and neutron matrix elements of the
electromagnetic current by
\bea
\langle X | J_Z^\mu\, | p \rangle
&=& (1 - 4 \sw) \langle X | J_\gamma^\mu\, | p \rangle\
             -\ \langle X | J_\gamma^\mu\, | n \rangle,
\label{eq:Icurrent}
\eea
neglecting the small contribution from strange quarks.
In general, the hadronic tensor can be decomposed in terms of
the $\gZ$ interference structure functions $\figz$ as
\bea
M W^{\mu\nu}_{\gZ}\
&=& - g^{\mu\nu} \fogz
    + {p^\mu p^\nu \over p\cdot q} \ftgz
    - i \epsilon^{\mu\nu\lambda\rho}
      {p_\lambda q_{\rho} \over 2 p \cdot q} \fthgz,
\label{eq:Had}
\eea
where $p$ is the four-momentum of the target hadron.
Note that the structure functions $\fogz$ and $\ftgz$ contribute
to the vector hadron contribution, while the $\fthgz$ structure
function appears only in the axial-vector hadron correction.
Combining Eqs.~(\ref{eq:Lep}) and (\ref{eq:Had}), the imaginary part
of the $\square_{\gZ}^V$ correction becomes \cite{Gorchtein:2008px,
Sibirtsev:2010zg, Arrington:2011dn}
\bea
\Im m\, \square_{\gZ}^V(E)
&=& {1 \over (s - M^2)^2}
    \int_{W_\pi^2}^s dW^2
    \int_0^{Q^2_{\rm max}} dQ^2\, {\alpha(Q^2) \over 1+Q^2/M_Z^2}
\nonumber\\
& & \hspace*{4.5cm}
\times
    \left[ \fogz
         + { s \left( Q^2_{\rm max}-Q^2 \right) \over
		      Q^2 \left( W^2 - M^2 + Q^2 \right) } \ftgz
    \right],
\label{eq:ImBoxV}					
\eea
where $s = M^2 + 2 M E$ is the total center of mass energy squared,
$W_{\pi}^2 = (M + m_\pi)^2$ is the mass at the pion threshold, and
$Q^2_{\rm max} = 2ME( 1 - W^2/s)$.  Following Ref.~\cite{Blunden:2011rd},
we include in Eq.~(\ref{eq:ImBoxV}) the $Q^2$ dependence in $\alpha(Q^2)$
arising from vacuum polarization contributions.

The most important inputs into Eq.~(\ref{eq:ImBoxV}) are the $\gZ$
interference structure functions $\figz$, which are functions of
two variables, usually taken to be $Q^2$ and the Bjorken scaling
variable $x = Q^2/2 p \cdot q$, or alternatively $Q^2$ and $W^2$.
Unfortunately, these functions are not well determined experimentally.
Although there are some data on $\fogz$ and $\ftgz$ at high $W$ and
$Q^2$, in the low-$W$ and $Q^2$ region, which is crucial to the
dispersion integrals, there is little or no information.
Unlike the electromagnetic structure functions, which can be fit
to the ample data available, the $\figz$ must be expressed
through models.  Given that it can be difficult to resolve the
accuracy of the models, the controversy in the literature over
the \regzv contribution is not surprising.

For later reference, we note here that the $F_1$ and $F_2$ structure
functions, for either $\gZ$ or electromagnetic ($\gg$) scattering,
can be related to the transverse ($\sigma_T$) and longitudinal
($\sigma_L$) electroweak boson production cross sections as
\begin{subequations}
\label{eq:sig_def}
\bea
F_1(W^2,Q^2)
&=& \left( \frac{W^2-M^2}{8 \pi^2 \alpha} \right)
    \sigma_T(W^2,Q^2),
\label{eq:f1gg} 		\\
F_2(W^2,Q^2)
&=& \left( \frac{W^2-M^2}{8 \pi^2 \alpha} \right)
    \frac{\nu}{M(1+\nu^2/Q^2)}
    \left[ \sigma_T(W^2,Q^2) + \sigma_L(W^2,Q^2) \right],
\label{eq:f2gg}
\eea
\end{subequations}
where $\nu = E - E'$ is the energy transfer.
For convenience one often defines the longitudinal structure function
as the combination of $F_1$ and $F_2$ structure functions given by
\bea
F_L &=&  \left( 1 + \frac{Q^2}{\nu^2} \right) F_2\ -\ 2 x F_1,
\label{eq:FL}
\eea
where the prefactor can also be written as $(1 + 4 x^2 M^2/Q^2)$.

\section{$\gZ$ interference structure functions}
\label{sec:SFs}

Most of the uncertainty in the calculation of the \bgZ correction
arises from the incomplete knowledge of the $\gZ$ structure functions.
There have been extractions of $\ftgz$ and $x\fthgz$ from
neutral current DIS by the H1 Collaboration at DESY \cite{Aaron:2012qi}
at very high $Q^2$ ($60 < Q^2 < 50,000$~GeV$^2$) and small $x$
($0.0008 < x < 0.65$) using longitudinally polarized lepton beams
at HERA.  However, these data have little overlap with the region
of most relevance for the dispersion integral, which receives
contributions primarily from high $x$ and low $Q^2$, where there
are no direct measurements.  Consequently, one must appeal to
models of the interference structure functions to estimate \bgZ.

In this section we review the models used in the literature for the
$\gZ$ structure functions, before presenting our constrained model,
which we refer to as the Adelaide-Jefferson Lab-Manitoba (AJM) model.
The construction of the models involves first choosing appropriate
electromagnetic structure functions $\figg$, and then transforming
these to the $\gZ$ case.  In describing the structure functions,
or equivalently the virtual boson-proton cross sections $\stl$ in
Eqs.~(\ref{eq:sig_def}), it is convenient to separate the full range of
kinematics into a resonance part and a smooth nonresonant background,
\bea
\sigma_{T,L} &=& \sigma_{T,L}^{\rm (res)} + \sigma_{T,L}^{\rm (bgd)}.
\label{eq:stll}
\eea
The $\sigma_{T,L}^{\rm (res)}$ term includes a sum over the prominent
low-lying resonances, while $\sigma_{T,L}^{\rm (bgd)}$ is determined
phenomenologically by fitting the inclusive scattering data
\cite{Christy:2007ve, Bosted:2007xd}.  Although such a separation
is inherently model dependent, as only the total cross section is
physical, it nevertheless provides a useful way to parametrize the
somewhat different behaviors of the cross sections in the low- and
high-$W$ regions.

For completeness, the following list summarizes the models for the
$\gZ$ structure functions that have been discussed in the literature:
\begin{enumerate}[label=(\roman{*})]
\itemsep0em
\item	color-dipole model \cite{Cvetic:2001ie, Cvetic:1999fi},
	referred to as ``Model I'' in Gorchtein {\it et al.} (GHRM)
	\cite{Gorchtein:2011mz};
\item	vector meson dominance (VMD) + Regge model
	\cite{Sakurai:1972wk, Alwall:2004wk}, referred to as
	``Model II'' by GHRM \cite{Gorchtein:2011mz};
\item	Sibirtsev {\it et al.} (SBMT) model \cite{Sibirtsev:2010zg},
	based on the Regge parametrization of Capella {\it et al.}
	\cite{Capella:1994cr};
\item	Carlson and Rislow (CR) model \cite{Carlson:2012yi,
	Rislow:2010vi}.
\end{enumerate}
The models \cite{Gorchtein:2011mz, Sibirtsev:2010zg, Rislow:2010vi,
Carlson:2012yi} differ primarily in the treatment of the background
contributions $\sigma_{T,L}^{\rm (bgd)}$ for the $\gZ$ interference,
the uncertainty on which is the main source of disagreement between
the various estimates of \bgZ.
For the resonance region, all of the models (with the exception
of SBMT \cite{Sibirtsev:2010zg}) use the Christy and Bosted (CB)
parametrization \cite{Christy:2007ve} of the electromagnetic structure
functions at low $W$, but differ in how these are transformed to the
$\gZ$ case.  Note, however, in both Model~I and Model~II of GHRM
some of the resonance parameters in the CB fit are modified to better
match the choice of background contribution \cite{Gorchtein:2011mz}.
In the following we discuss both the resonance and background content
of these models in more detail.

\subsection{Resonances}

The CB parametrization \cite{Christy:2007ve} of $F_{1,2}^{\gg}$
fits the resonance region electron-proton scattering data in terms
of the seven most important resonances
($P_{33}(1232)$,
 $P_{11}(1440)$, 
 $D_{13}(1520)$,
 $S_{11}(1535)$,
 $S_{15}(1650)$,
 $F_{15}(1680)$ and an $l=3$ state with mass 1934~MeV),
and generally agrees with the data to within $5\%$.
The CB fit is used as the basis for the resonance models of
Carlson and Rislow \cite{Carlson:2012yi}, and Gorchtein {\it et al.}
\cite{Gorchtein:2011mz}, with the latter using slightly modified
parameters for $\sigma_{T,L}^{\rm (res)}$ in their Models~I and II.
Sibirtsev {\it et al.} \cite{Sibirtsev:2010zg}, on the other hand,
perform their own fit of the data, incorporating the four resonances
 $P_{33}(1232)$,
 $D_{13}(1520)$,
 $F_{15}(1680)$ and
 $F_{37}(1950)$,
and also obtain a reasonably good description of the data.

Modifying the electromagnetic structure functions to obtain their
interference analogs involves modifying the contribution from each
resonance $R$ by a ratio that takes into account the differences
between the electromagnetic and weak neutral transition amplitudes,
according to Eq.~(\ref{eq:Icurrent}).  For the transverse cross
section GHRM define this ratio for a proton as \cite{Gorchtein:2011mz}
\bea
\xi_R
&\equiv& \frac{\sigma_{T,R}^{\gZ}}{\sigma_{T,R}^{\gg}}\
 =\ (1 - 4\sin^2 \theta_W) - y_R,
\label{eq:xiR}
\eea
where
\begin{equation}
y_R = \frac{ A_{R, \frac{1}{2}}^p \, A_{R, \frac{1}{2}}^{n^*}
	   + A_{R, \frac{3}{2}}^p \, A_{R, \frac{3}{2}}^{n^*} }
	   { \big| A_{R, \frac{1}{2}}^p \big|^2
	   + \big| A_{R, \frac{3}{2}}^p \big|^2 },
\label{eq:yr}
\end{equation}
with $A_{R, \lambda }^N$ the transition amplitude from a proton or
neutron to a resonance $R$ with helicity $\lambda = \frac{1}{2}$
or $\frac{3}{2}$.  The amplitudes $A_{R, \lambda }^N$ are assumed
by GHRM to be $Q^2$ independent, and their values determined from
electromagnetic decays at $Q^2=0$ \cite{Nakamura:2010zzi}.
The ratio for the longitudinal cross section is taken to be
equal to the transverse ratio in both Models~I and II of GHRM.

Carlson and Rislow \cite{Carlson:2012yi} use a similar ratio to that
in Eq.~(\ref{eq:xiR}) (which they label as $C_R$), but include in
addition a $Q^2$ dependence in the amplitudes derived from the MAID
unitary isobar model \cite{Tiator:2011pw}.  For comparison, CR also
calculate the transition amplitudes using a constituent quark model
\cite{Rislow:2010vi}.

Finally, Sibirtsev {\it et al.} \cite{Sibirtsev:2010zg} use the
conservation of the vector current and isospin symmetry to set
the ratio for isospin-3/2 states to $(1+Q_W^p) \approx (2-4\sw)$.
For the isospin-1/2 resonances, such as the $D_{13}(1520)$, SU(6)
quark model wave functions are used to estimate the ratio of couplings.
The similarity of the magnitudes of the weak and electromagnetic
couplings was used by SBMT to justify approximating the ratio
$\xi_R$ by 1.

\subsection{Background}

\subsubsection{Electromagnetic structure functions}

Although the CB parametrization \cite{Christy:2007ve} includes
a background $\sigma_{T,L}^{\rm (bgd)}$ at low $W$ ($W < 3$~GeV),
to describe the nonresonant contributions to the electromagnetic
structure functions at $W > 3$~GeV requires a model for the
background which is also valid at large $W$.
In the calculation of GHRM \cite{Gorchtein:2011mz}, the color dipole
model from Cvetic {\it et al.} \cite{Cvetic:2001ie, Cvetic:1999fi}
is used for Model~I, while the VMD+Regge model of Alwall and Ingelman
\cite{Alwall:2004wk} is employed for Model~II.  Since the latter was
shown by GHRM to introduce the largest uncertainty in \bgZ, it will
be the main focus of our attention.

According to the VMD hypothesis, the interaction of a photon $\gamma$
with a hadron proceeds through transitions to vector mesons $V$ (with
$V = \rho$, $\omega$ or $\phi$), with strength $\sqrt{4\pi\alpha}/f_V$,
where $f_V$ is the electromagnetic decay constant of $V$.
The three vector mesons saturate around 80\% of the total
photoproduction cross section \cite{Sakurai:1972wk}.
The remainder is usually attributed to contributions from higher
masses, which are modeled by a continuum of states starting at
mass $m_0 \approx 1.4$~GeV \cite{Sakurai:1972wk}.
(In the case of the color dipole model \cite{Cvetic:2001ie,
Cvetic:1999fi, Kuroda:2011dw}, the photon is assumed to interact
with the hadron through coupling to uncorrelated $q \bar{q}$ states
instead of mesons.)
Following Ref.~\cite{Alwall:2004wk}, we neglect the off-diagonal
terms in the mass integral, which is known to be a good approximation
for scattering from nucleons \cite{Fraas:1974gh}.  The transverse and
longitudinal virtual photon-nucleon cross sections can then be
expressed as \cite{Alwall:2004wk}
\begin{subequations}
\bea
\sigma_T^{\rm VMD}
&=& \sigma_{\gamma N}
    \Bigg[ \sum_V\, r_V \frac{1}{(1 + Q^2/m_V^2)^2}\
	+\ r_C\, \frac{1}{1 + Q^2/m_0^2}
    \Bigg],
\label{eq:sTv} \\
\sigma_L^{\rm VMD}
&=& \sigma_{\gamma N}
    \Bigg[ \sum_V\, r_V\, \xi_V\,
	\frac{Q^2/m_V^2}{(1 + Q^2/m_V^2)^2}		\nn\\
&& \hspace*{1cm}
     +\ r_C\, \xi_C
	\left( \frac{m_0^2}{Q^2} \ln(1+Q^2/m_0^2)
	     - \frac{1}{1 + Q^2/m_0^2}
	\right)
    \Bigg],
\label{eq:sLv}
\eea
\end{subequations}
where $\sigma_{\gamma N}$ is the real photon-nucleon cross section,
and the constants $r_V \sim 1/f_V^2$ represent the relative
contributions from the individual vector mesons $V$, with
$r_C = 1 - \sum\limits_V r_V$ being the continuum fraction
\cite{Alwall:2004wk}.  Phenomenologically, the $r_V$ values are
determined as $r_V = \left\{ 0.67,\, 0.062,\, 0.059 \right\}$ for
$V = \rho$, $\omega$ and $\phi$, respectively \cite{Bauer:1977iq}.
As we shall see below, $r_C$ plays a critical role in determining
the uncertainty on the interference cross sections.
The parameters $\xi_V$ and $\xi_C$ allow for different behavior of
the transverse and longitudinal components of the vector mesons,
although in practice these are usually set equal, $\xi_V = \xi_C$,
in order to fit the available data.
Note that despite the apparent $1/Q^2$ dependence in the second
term of $\sigma_L^{\rm VMD}$ in Eq.~(\ref{eq:sLv}), one can verify
by expanding the logarithm for small $Q^2$ that the longitudinal
cross section does in fact vanish in the $Q^2 \to 0$ limit.
According to Regge theory, the real photon cross section can be
parametrized as a sum of two terms \cite{Donnachie:1992ny},
\bea
\sigma_{\gamma N}
&=& A_\gamma\, s_\gamma^\epsilon + B_\gamma\, s_\gamma^{-\eta},
\eea
where $s_\gamma \equiv W^2$, with the exponents $\epsilon$ and $\eta$
giving the energy dependence of the Pomeron and Reggeon terms,
which have coefficients $A_\gamma$ and $B_\gamma$, respectively.

In the model of SBMT, the background is parametrized according to the
structure function fit of Capella {\it et al.} \cite{Capella:1994cr},
with several parameters adjusted to better describe recent data, as
discussed in Ref.~\cite{Sibirtsev:2010zg}.  The parametrization of the
$\ftgg$ structure function, which is valid for all $Q^2$, is again
given by a sum of Pomeron ($P$) and Reggeon $R$ exchange terms,
\bea
\hspace*{-0.4cm}
\ftgg(x,Q^2)
&=& A_{\rm P}\, x^{-\Delta} (1-x)^{n+4}
    \left[ \frac{Q^2}{Q^2 + \Lambda_{\rm P}^2} \right]^{1+\Delta}\
 +\ A_{\rm R}\, x^{1-\alpha_{\rm R}} (1-x)^n
    \left[ \frac{Q^2}{Q^2 + \Lambda_{\rm R}^2} \right]^{\alpha_{\rm R}},
						\nn\\
& &
\eea
where $\Delta$ and $n$ are both functions of $Q^2$, and
$A_{\rm P}$, $\Lambda_{\rm P}$, $A_{\rm R}$, $\Lambda_{\rm R}$
and $\alpha_{\rm R}$ are fit parameters \cite{Capella:1994cr}.
The $\fogg$ structure function is obtained by SBMT from a
parametrization of the ratio of longitudinal to transverse cross
sections.  From Eqs.~(\ref{eq:sig_def}) this can be written as
\bea
\frac{\sigma_L}{\sigma_T}
&=& \left( 1 + \frac{4 M^2 x^2}{Q^2} \right) \frac{F_2}{2 x F_1} - 1,
\label{eq:LTdef}
\eea
which is parametrized by a sum of exponentials \cite{Sibirtsev:2010zg}.

While the above models use the same background parametrization over the
entire range of kinematics, CR \cite{Rislow:2010vi, Carlson:2012yi} on
the other hand divide their dispersion integral into three distinct
regions, each described by a different model.
In particular, the resonance region at low $W$ is described in terms of
the CB fit to $\sigma_{T,L}^{\rm (res)}$ and $\sigma_{T,L}^{\rm (bgd)}$
\cite{Christy:2007ve}, while for the high-$W$, low-$Q^2$ region,
CR use the Capella {\it et al.} structure function parametrization.
For high $W$ and high $Q^2$, a partonic description is employed using
the CT10 global fit \cite{Lai:2010vv} of parton distribution functions
(PDFs).

\subsubsection{$\gZ$ structure functions}

To construct the nonresonant background contributions to the
transverse and longitudinal $\gZ$ cross sections, the electromagnetic
cross sections need to be rescaled by the ratio
	$\sigma_{T, L}^{\gZ} / \sigma_{T, L}^{\gg}$,
as for the resonance components.
For Model~II of GHRM \cite{Gorchtein:2011mz}, a generalization of
the VMD model is used, assuming the $\gZ$ cross section for vector
meson $V$ is given by the analogous $\gg$ cross section scaled by
the ratio $\kappa_V$ of weak and electric charges,
\bea
\sigma_{T, L}^{\gZ (V)} &=& \kappa_V\, \sigma_{T, L}^{\gg (V)},
\eea
where
\begin{subequations}
\bea
\kappa_\rho   &=& 2 - 4 \sw,	\\
\kappa_\omega &=& -4 \sw,	\\
\kappa_\phi   &=& 3 - 4 \sw
\eea
\end{subequations}
correspond to the isovector, isoscalar and strange quark
components of the electroweak current, respectively.
This allows the ratio of $\gZ$ to $\gg$ cross sections to
be written as \cite{Gorchtein:2011mz}
\be
\frac{\sigma_{T, L}^{\gZ}}
     {\sigma_{T, L}^{\gg}}
= \frac{ \kappa_\rho + \kappa_\omega\, R_{\omega}^{T, L}(Q^2)
		     + \kappa_\phi\, R_{\phi}^{T, L}(Q^2)
		     + \kappa_C^{T, L}\, R_C^{T, L}(Q^2)}
       {1 + R_{\omega}^{T, L}(Q^2)
	  + R_{\phi}^{T, L}(Q^2)
	  + R_C^{T,L}(Q^2)},
\label{eq:GHRM45}
\ee
where $R_V^{T, L}$ is the ratio of cross sections for $V$
and the $\rho$ meson,
\bea
R_V^{T, L}
&\equiv& \frac{ \sigma_{T,L}^{\gg (V)} }
	 { \sigma_{T,L}^{\gg (\rho)} }\
 =\ \frac{f_\rho^2}{f_V^2}
    \left( \frac{1 + Q^2/m_\rho^2}{1 + Q^2/m_V^2} \right)^2.
\eea
The corresponding ratio $R_C^{T, L}$ of the continuum to $\rho$
contributions is given by
\begin{subequations}
\bea
R_C^T
&=& \frac{r_C}{r_\rho}
    \left( \frac{1 + Q^2/m_\rho^2}{1 + Q^2/m_0^2} \right)^2,
\label{eq:ConRatT}					\\
R_C^L
&=& \frac{r_C}{r_\rho}
    \left[ \frac{m_0^2}{Q^2} \ln(1 + Q^2/m_0^2)\
	-\ \frac{1}{1 + Q^2/m_0^2}
    \right]
\Big/
    \left[ \frac{Q^2/m_\rho^2}{(1 + Q^2/m_\rho^2)^2}
    \right],
\label{eq:ConRatL}
\eea
\end{subequations}
with the continuum mass parameter set to $m_0 = 1.5$~GeV
\cite{Gorchtein:2011mz}.
The parameters $\kappa_C^{T, L}$ in Eq.~(\ref{eq:GHRM45}) denote
the ratios of the $\gZ$ and $\gg$ continuum contributions to the
cross section.  Unlike for the discrete vector meson terms, the VMD
model does not prescribe a simple charge ratio factor to modify the
continuum part of the cross section.  In view of this, GHRM proceed
by assigning a 100\% uncertainty on this contribution.  As we will
see below, this assumption gives the largest contribution to the
uncertainty on \bgZ.

For Model~I of GHRM, the same general form for the $\gZ$ cross sections
is used as in Eq.~(\ref{eq:GHRM45}), but with different individual
contributions $R_V^{T,L}$.  Whereas in Model~II, the $R_V^{T,L}$ are
functions of $Q^2$, in Model~I these become constants with relative
strengths determined by squares of quark electric charges, with the
continuum contribution associated with the $J/\psi$ meson
\cite{Gorchtein:2011mz},
$\{ \rho : \omega : \phi : J/\psi \} =
 \{ 1 : 1/9 : 2/9 : 8/9 \}$.
%
%
Similarly, a 100\% uncertainty is assumed for the $J/\psi$
term in Model~I.

In the SBMT model \cite{Sibirtsev:2010zg}, the $\gZ$ structure functions
at low $x$ are approximated by their electromagnetic counterparts.
This is motivated by the approximate flavor independence of sea quark
distributions in the low-$x$ region, and the similarity of the sum
of the electroweak couplings for three quark flavors,
  $\left(\sum_q e_q\, g_V^q\right) / \left(\sum_q e_q^2\right)
   = 2 - 4 \sw \approx 1$ \cite{Gorchtein:2008px},
where $e_q$ and $g_V^q$ are the electric and weak vector charges
of quark $q$, respectively.
At larger $x$ \mbox{($x \gtrsim 0.4$)}, however, SBMT compute $\figz$
using a ratio of leading twist (LT) structure functions computed
from the Martin-Roberts-Stirling-Thorne parton distribution functions \cite{Martin:2002aw},
\bea
\figz &=& \left( \frac{\figz}{\figg} \right)^{\rm LT} \figg.
\eea
At these $x$ values, SBMT note that the flavor dependence of
the parton distributions renders the interference function
approximately $30\%\!-\!40\%$ smaller than the electromagnetic one. The
functions $\figg$ therefore provide an upper limit on $\figz$.

Finally, for the CR model \cite{Carlson:2012yi, Rislow:2010vi}
the method for modifying the $\gg$ background cross sections
depends on the kinematic region of $W$ and $Q^2$.
In the resonance region, CR take the average of the high energy
($x \to 0$) limit ($u=d=s$), in which
  $\figz/\figg = 2-4\sw$,
and the SU(6) quark limit ($u=2d, s=0$), in which
  $\figz/\figg = 5/3-4\sw$,
to convert the electromagnetic background from the CB structure
function parametrization \cite{Christy:2007ve}.
For the low-$Q^2$, high-$W$ region, CR apply the same ratio to the
Capella {\it et al.} \cite{Capella:1994cr} parametrization as SBMT,
while in the DIS region they compute the $\figz$ structure functions
directly from LT parton distributions \cite{Lai:2010vv}.

Using these models for the resonance and nonresonant background
contributions to the $\gZ$ structure functions, the analyses of
GHRM \cite{Gorchtein:2011mz}, SBMT \cite{Sibirtsev:2010zg} and
CR \cite{Rislow:2010vi} estimate the \bgZ correction at the
\qwe energy to be
\begin{subequations}
\bea
\Re e\, \square_{\gZ}^V
&=& (5.4 \pm 2.0) \times 10^{-3}
	\hspace*{2cm} {\rm [GHRM]} \\
\Re e\, \square_{\gZ}^V
&=& (4.7_{\ -0.4\, }^{\ +1.1\, }) \times 10^{-3}
	\hspace*{2.25cm} {\rm [SBMT]} \\
\Re e\, \square_{\gZ}^V
&=& (5.7 \pm 0.9) \times 10^{-3}
	\hspace*{2cm} {\rm [CR]}
\eea
\label{eq:ReBox_other}
\end{subequations}
respectively.
The GHRM result for the central value of \regzv is the average of
Models~I and II, but with the dominant background error taken from
the larger of the two, in this case Model~II.
The GHRM analysis also estimates the effect of the $t$ dependence
of the \bgZ correction, from $t=0$ in the dispersion formalism to
$t = -0.03$~GeV$^2$ in the \qwe experiment, finding a decrease of
approximately 1.3\%, with a similar uncertainty on the correction
at the \qwe point.

The central values of all the calculations agree within the quoted
uncertainties; however, the error on the GHRM value is twice as
large as those on the SBMT and CR calculations, even though the
SBMT estimate includes a fairly conservative uncertainty on the
input $\gg$ structure functions.
Given the importance of the \bgZ correction to the extraction of
the weak mixing angle from the \qwe measurement, it is vital that
the origin of this difference be understood, and ways of further
reducing the uncertainty explored.

\subsection{Adelaide-Jefferson Lab-Manitoba model}

To proceed with our analysis of the $\gZ$ correction, we define here
the ingredients of our AJM
model, within which we will study in detail the various contributions
to \regzv and their uncertainties.  We draw on the valuable
experience obtained with the existing models \cite{Gorchtein:2008px,
Gorchtein:2011mz, Sibirtsev:2010zg, Rislow:2010vi, Carlson:2012yi},
and incorporate into the AJM model some of the more robust features
of the previous analyses.
Most importantly, we consider additional constraints from existing
data on some of the model parameters which were unconstrained in
the earlier work.  We will find that indeed data on PDFs near the
resonance-DIS transition, together with new results on
inclusive parity-violating electron scattering asymmetries,
place significant constraints on the models, in particular on
the background contribution.

\onecolumngrid
\begin{figure*}
\includegraphics[width=3.5in]{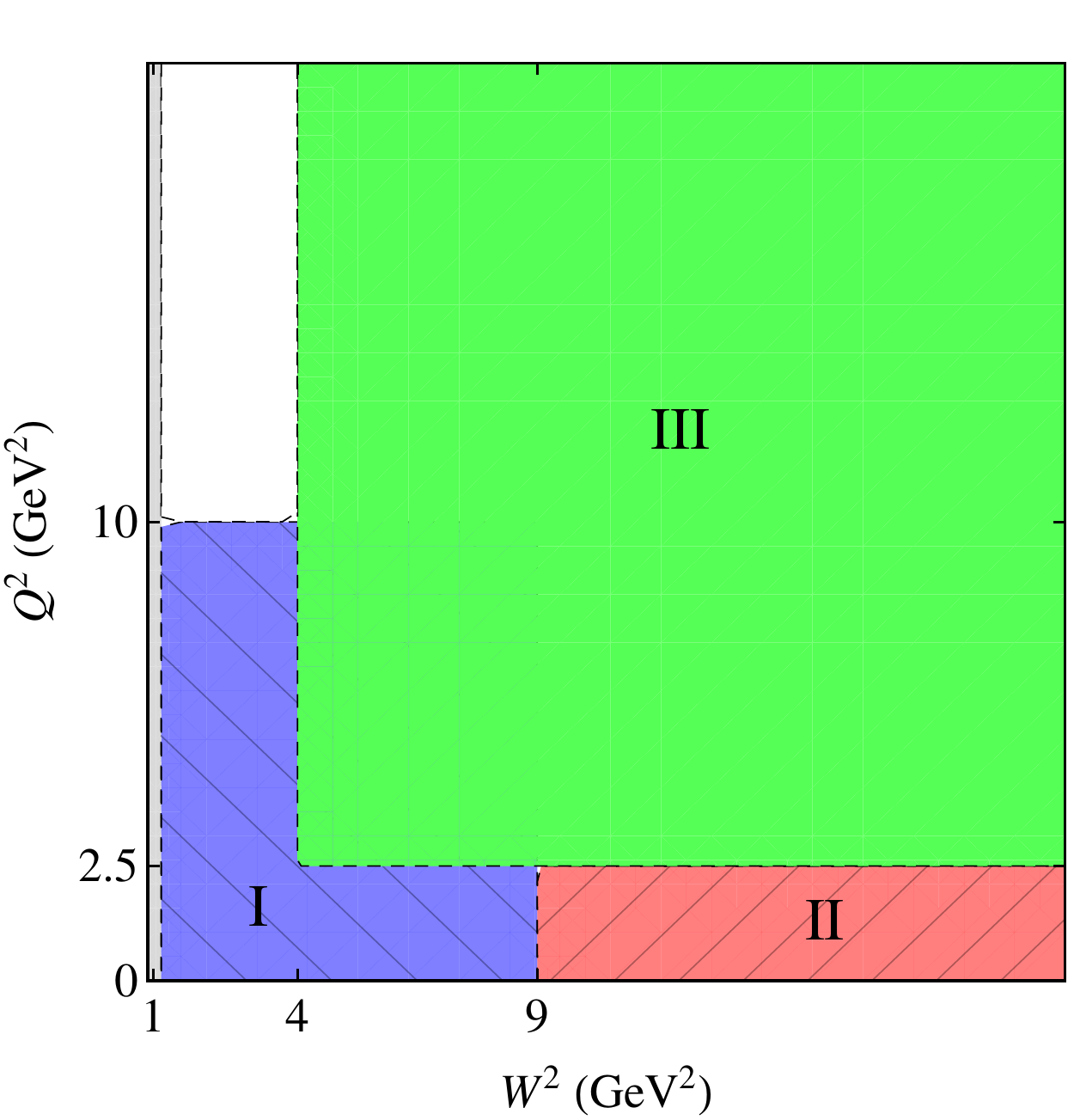}
\caption{(color online)
	Kinematic regions contributing to the \bgZv integrals
	in the AJM model:
	Region~I (blue shaded) at low $W$ and low $Q^2$ is described
	by the CB $F_{1,2}^{\gg}$ fit \cite{Christy:2007ve},
	transformed to the $\gZ$ case;
	Region~II (red shaded) represents the high-$W$, low-$Q^2$
	domain as in Ref.~\cite{Alwall:2004wk} (or the GHRM Model~II
	\cite{Gorchtein:2011mz}), transformed to $\gZ$;
	and Region~III (green shaded) at high $W$ and high $Q^2$
	is described by global PDF fits to high-energy scattering
	data \cite{Alekhin:2012ig}.}
\label{fig:W2Q2}
\end{figure*}

\subsubsection{$\gg$ structure functions}

Following CR \cite{Rislow:2010vi, Carlson:2012yi}, we divide the
integrals in Eq.~(\ref{eq:ImBoxV}) into distinct regions of $W^2$
and $Q^2$, using specific models to parametrize the $\gZ$ structure
functions in each region.  This is illustrated in Fig.~\ref{fig:W2Q2},
where the $W^2$ and $Q^2$ divisions and the models describing them
are indicated.  Although the boundaries between the regions are
clearly defined, the models themselves overlap, allowing important
checks to be made on the continuity of the descriptions across the
boundaries.

\begin{figure}[t]
\includegraphics[width=\textwidth]{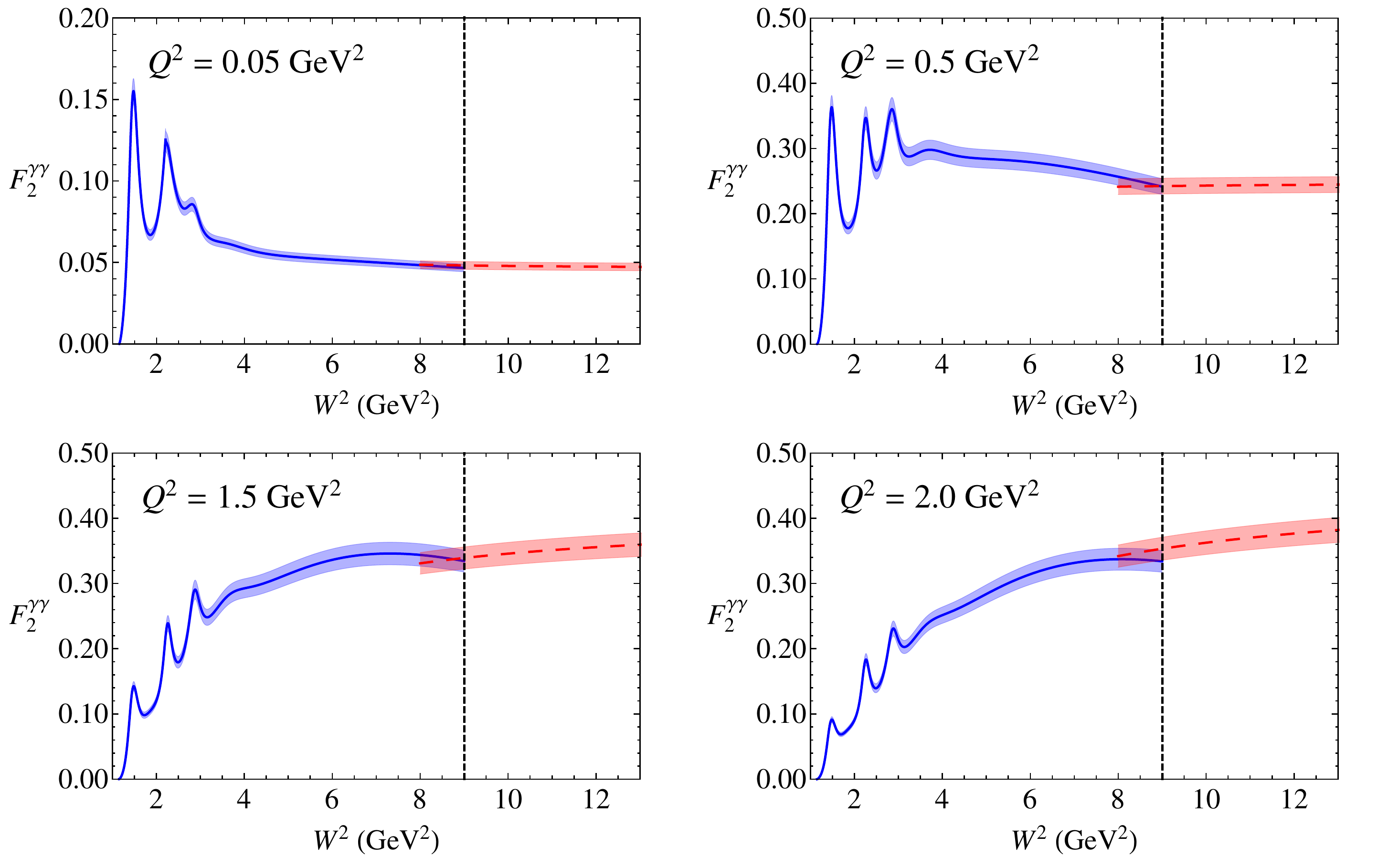}
\caption{(color online)
	Proton $\ftgg$ structure function versus $W^2$ at fixed
	$Q^2=0.05$, 0.5, 1.5 and 2~GeV$^2$ for the CB fit
	\cite{Christy:2007ve} at low $W$ (blue solid) and VMD+Regge
	parametrization \cite{Alwall:2004wk} at high $W$ (red dashed).
	The boundary between these (corresponding to Regions~I and II
	in Fig.~\ref{fig:W2Q2}) is indicated by the vertical dashed
	line at $W^2 = 9$~GeV$^2$.}
\label{fig:F2gCV}
\end{figure}

For the input $\gg$ structure functions, we use the CB parametrization
\cite{Christy:2007ve} to describe the low-$W$ region (Region~I) at
$W_\pi < W < 2$~GeV for all $Q^2$ up to 10~GeV$^2$.
In fact, the strong suppression of the resonance transition form
factors with increasing $Q^2$ results in negligible resonance
contributions already beyond $Q^2 \approx 2$~GeV$^2$.
Since the CB fit also describes data up to $W^2 = 9$~GeV$^2$,
we use it in the higher-$W$ region for $Q^2 < 2.5$~GeV$^2$,
as indicated by the blue shaded area in Fig.~\ref{fig:W2Q2}.

\begin{figure}[t]
\includegraphics[width=\textwidth]{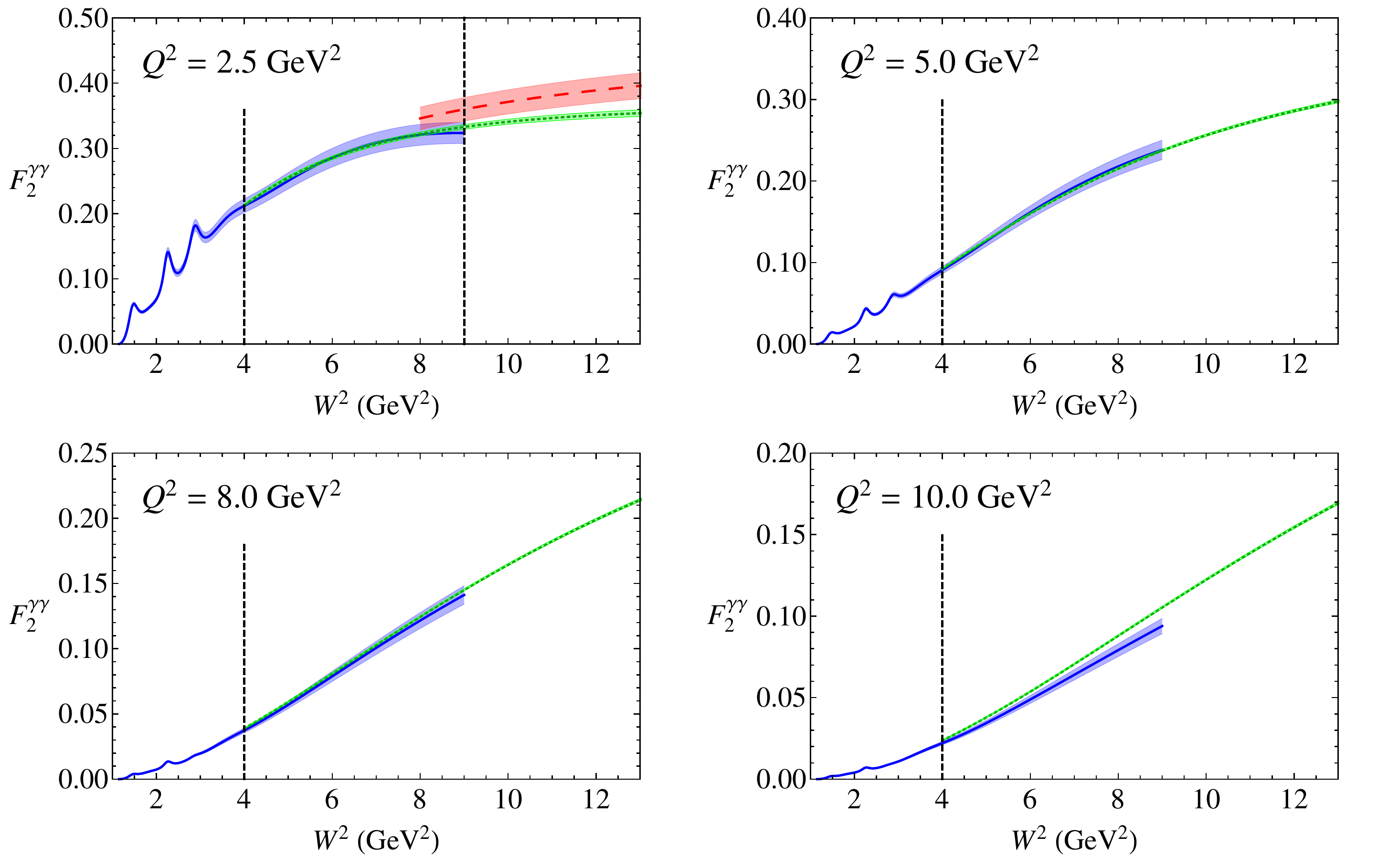}
\caption{(color online)
	Proton $\ftgg$ structure function versus $W^2$ at
	fixed $Q^2=2.5$, 5, 8 and 10~GeV$^2$ for the CB fit
	\cite{Christy:2007ve} at low $W$ (blue solid) and the
	ABM11 PDF parametrization \cite{Alekhin:2012ig}
	at high $W$ (green dotted), with the boundary between
	Regions~I and III at $W^2=4$~GeV$^2$ indicated by the
	vertical line.
	For the $Q^2=2.5$~GeV$^2$ panel, the matching with
	the VMD+Regge model \cite{Alwall:2004wk} (red dashed),
	corresponding to the boundary between Regions~I and II,
	is indicated by the vertical line at $W^2=9$~GeV$^2$.}
\label{fig:F2gCP}
\end{figure}

At higher $W$, corresponding to kinematics where Regge theory
is applicable, the VMD+Regge model of Alwall and Ingelman
\cite{Alwall:2004wk} is combined with a modified CB resonance
contribution ({\it cf.} Table~II of Ref.~\cite{Gorchtein:2011mz})
to describe the structure functions for $W^2 > 9$~GeV$^2$ and
$Q^2 < 2.5$~GeV$^2$ (Region~II, red shaded area in Fig.~\ref{fig:W2Q2}).
Of course, at these values of $W$ the resonances will contribute
very little to the dispersion integral in Eq.~(\ref{eq:ImBoxV}),
which will be contaminated by the background contribution.
This model also forms the basis for Model~II of GHRM
\cite{Gorchtein:2011mz}.
The matching of the CB and VMD+Regge parametrizations at the
boundary between the low-$W$ and high-$W$ regions is illustrated
in Fig.~\ref{fig:F2gCV} for the $\ftgg$ structure function as
a function of $W^2$, at several fixed values of $Q^2$,
from $Q^2 = 0.05$ to 2~GeV$^2$.  The agreement between the
two models in the region of overlap is clearly excellent.
For the structure function in the VMD+Regge model, we have
assumed a conservative 5\% uncertainty, similar to that for
the CB parametrization.

In the DIS region at high $W$ and high $Q^2$ (green shaded area in
Fig.~\ref{fig:W2Q2}), the structure functions can be computed in
terms of global PDFs, for which we use the next-to-next-to-leading
order (NNLO) fit by Alekhin {\it et al.} (ABM11) \cite{Alekhin:2012ig}.
This fit includes both leading twist and higher twist contributions,
allowing for descriptions of data for $Q^2 > 2.5$~GeV$^2$ and
$W > 1.8$~GeV, which overlaps partially with the CB \cite{Christy:2007ve}
and VMD+Regge \cite{Alwall:2004wk} parametrizations.
(Other similar global fits, such as those in Refs.~\cite{Accardi:2011fa, 
Owens:2012bv, Ball:2010de, JimenezDelgado:2009tv, Martin:2009iq}, give
very similar results, and differences between the parametrization
generally lie within the PDF uncertainties.)
The transition between DIS kinematics (Region~III) and the models
describing the lower-$W$ and $Q^2$ regions is illustrated in
Fig.~\ref{fig:F2gCP} for $\ftgg$ at $Q^2 = 2.5$~GeV$^2$
(where the transitions between all three parametrizations
are shown at $W^2=9$~GeV$^2$) and at higher $Q^2$ values, up to
$Q^2 = 10$~GeV$^2$, for the transition between Regions~I and III.
Again, the models generally match very well across these kinematic
boundaries.

\begin{figure}[t]
\includegraphics[width=\textwidth]{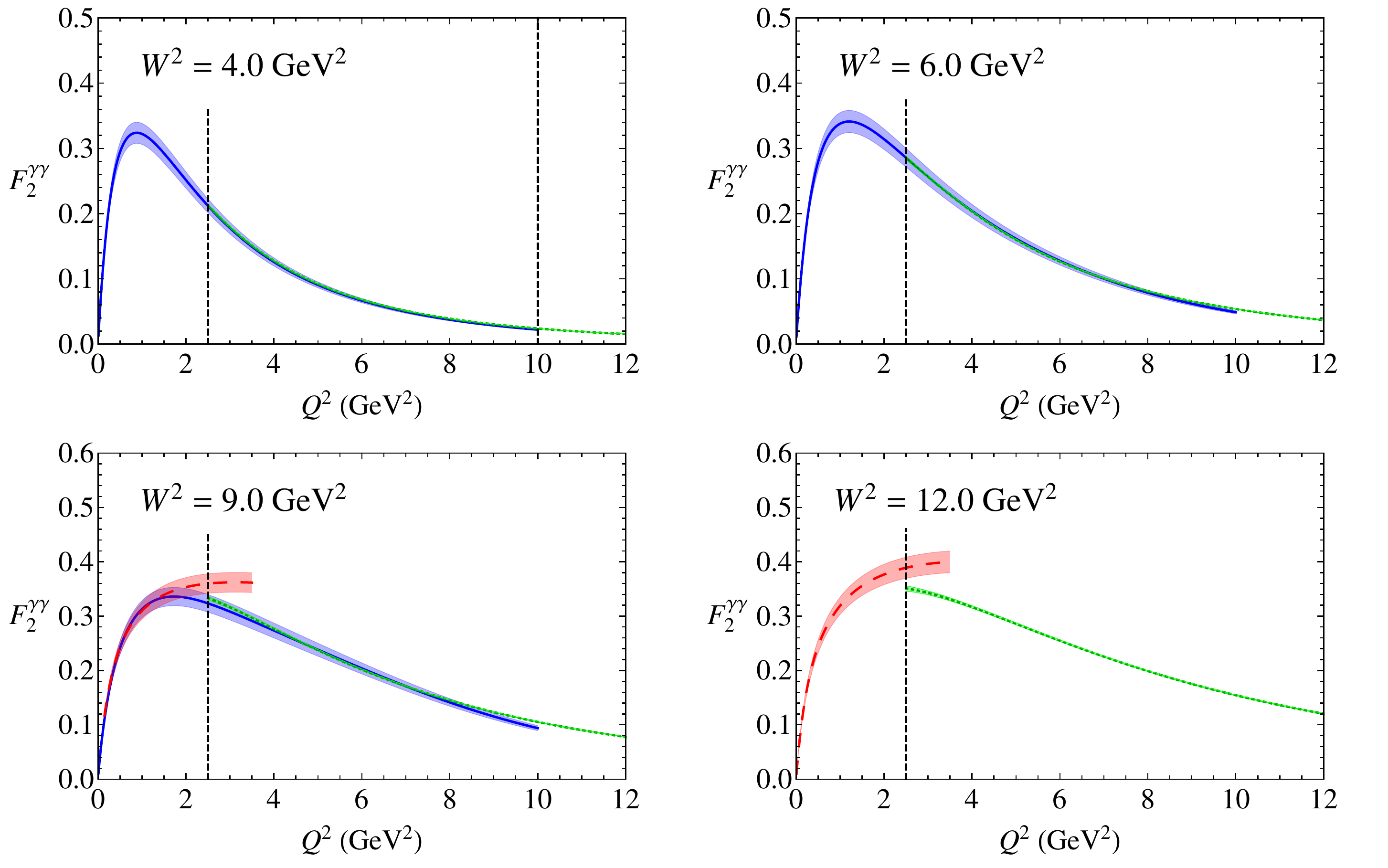}
\caption{(color online)
	Proton $\ftgg$ structure function versus $Q^2$ at
	fixed $W^2=4$, 6, 9 and 12~GeV$^2$ for the CB fit
	\cite{Christy:2007ve} (blue solid), the ABM11 PDF
	parametrization \cite{Alekhin:2012ig} (green dotted),
	and the VMD+Regge model \cite{Alwall:2004wk} (red dashed),
	with the boundaries between Regions~I, II and III
	indicated by the vertical lines at fixed $Q^2$.
	Note that the small disagreement between the VMD+Regge
	model and the PDF parametrization for $Q^2=2.5$ GeV$^2$
	appears only at larger $W^2$ values where the contribution
	to the dispersion integral is small.}
\label{fig:F2gVP}
\end{figure}

The boundaries between the three regions can also be displayed for fixed
$W^2$ as a function of $Q^2$, as illustrated in Fig.~\ref{fig:F2gVP}.
The matching of Regions~I and II for $W^2=4$~GeV$^2$ shows excellent
agreement between the CB \cite{Christy:2007ve} and ABM11 PDF
\cite{Alekhin:2012ig} parametrizations at $Q^2=2.5$~GeV$^2$.
At the highest $W$ value at which the CB fit is valid, $W^2=9$~GeV$^2$,
the agreement between the models describing all three regions is also
quite good.  For larger $W$ ($W^2 \gtrsim 10$~GeV$^2$) the VMD+Regge
model \cite{Alwall:2004wk} slightly exceeds the PDF parametrization.
However, this generally occurs at the edge of the kinematic boundary
between Regions~II and III, where the contribution to the imaginary
part of \bgZ in Eq.~(\ref{eq:ImBoxV}) is very small.

\subsubsection{$\gZ$ structure functions}

Having detailed the forms of the electromagnetic structure functions,
we now turn to their $\gZ$ interference analogs.
For the low-$W$/low-$Q^2$ region dominated by the nucleon resonances,
the transverse and longitudinal $\gg$ cross sections parametrized in
the CB fit \cite{Christy:2007ve} are modified using the ratio $\xi_R$
in Eq.~(\ref{eq:xiR}), with the parameter $y_R$ determined from the
proton and neutron helicity amplitudes as in Eq.~(\ref{eq:yr}).
This follows closely the approach adopted by GHRM
\cite{Gorchtein:2011mz}, but, importantly, differs in the way the
uncertainties on the helicity amplitudes $A_{R,\lambda}^N$ are
determined.

In particular, GHRM combined the uncertainties on the amplitudes
by adding extremal values of each, which implicitly assumes a
uniform error distribution rather than the standard Gaussian one.
Adding errors linearly clearly overestimates the uncertainties,
and in the AJM analysis we adopt the more conventional Gaussian
distribution to add the errors in quadrature.
(When combining all of the uncertainties on the final \regzv
value, however, GHRM add the errors in quadrature.)
In Table~\ref{tab:yr} the $y_R$ values for the proton and their
uncertainties computed using both methods are shown for comparison.
For completeness, we also list the $y_R$ values for the neutron
and deuteron, with uncertainties added in quadrature, which will
be needed in subsequent sections.
For the isospin-$\frac32$ $P_{33}(1232)$ and $F_{37}(1950)$
resonances, the uncertainties on the helicity amplitudes are given
by the Particle Data Group (PDG) \cite{Beringer:2012zz} as zero.
To be conservative, however, we follow GHRM \cite{Gorchtein:2011mz}
and include a 10\% uncertainty on the $P_{33}(1232)$ and a 100\%
uncertainty on the $F_{37}(1950)$ resonance \cite{Christy:2007ve,
Bosted:2007xd}.

\begin{savenotes}
\begin{table}[t]
\begin{center}
\caption{Electromagnetic to $\gZ$ resonance cross section
	transformation ratios $y_R$ from Eq.~(\ref{eq:yr})
	for the proton, neutron and deuteron in the AJM model,
	compared with the proton ratio in the GHRM model 
	\cite{Gorchtein:2011mz}.  The AJM model values in
	parentheses use helicity amplitudes from the earlier
	2010 PDG \cite{Nakamura:2010zzi}, as utilized by GHRM.
	The errors labeled with the asterisks $(*)$ are values
	corrected \cite{Gorchtein:priv} from those in
	Ref.~\cite{Gorchtein:2011mz}.}
\resizebox{\textwidth}{!}{%
\begin{tabular}{ c c | c c c c c c c c c c c c c }	\hline
\hspace{0.15in} & \hspace{0.35in}
	& \hspace{0.1in}   $P_{33}(1232)$
	& \hspace{0.1in} & $P_{11}(1440)$
	& \hspace{0.1in} & $D_{13}(1520)$
	& \hspace{0.1in} & $S_{11}(1535)$
	& \hspace{0.1in} & $S_{11}(1665)$
	& \hspace{0.1in} & $F_{15}(1680)$
	& \hspace{0.1in} & $F_{37}(1950)$		\\ \hline
\hspace{0.0in} & $p$ (AJM)
	& \hspace{0.1in}   $-1.0 \pm 0.1$
	& \hspace{0.1in} & $-0.67 \pm 0.17$
	& \hspace{0.1in} & $-0.84 \pm 0.17$
	& \hspace{0.1in} & $-0.51 \pm 0.35$
	& \hspace{0.1in} & $-0.28 \pm 0.41$
	& \hspace{0.1in} & $-0.27 \pm 0.08$
	& \hspace{0.1in} & $-1 \pm 1$			\\
\hspace{0.0in} &
	& \hspace{0.1in}
	& \hspace{0.1in} & $(-0.62 \pm 0.16)$
	& \hspace{0.1in} & $(-0.77 \pm 0.08)$
	& \hspace{0.1in} &
	& \hspace{0.1in} &
	& \hspace{0.1in} &
	& \hspace{0.1in} &				\\
\hspace{0.0in} & $p$ (GHRM)
	& \hspace{0.1in}   $-1.0 \pm 0.1$
	& \hspace{0.1in} & $-0.62^{+0.19}_{-0.20}$
	& \hspace{0.1in} & $-0.77^{+0.122}_{-0.125}${}$^{\ (*)}$
	& \hspace{0.1in} & $-0.51^{+0.35}_{-0.71}$
	& \hspace{0.1in} & $-0.28^{+0.45}_{-0.69}${}$^{\ (*)}$
	& \hspace{0.1in} & $-0.27^{+0.10}_{-0.12}$
	& \hspace{0.1in} & $-1 \pm 1$			\\ \hline
\hspace{0.in} & $n$ (AJM)
	& \hspace{0.1in}   $-1.0 \pm 0.1$
	& \hspace{0.1in} & $-1.50 \pm 0.39$
	& \hspace{0.1in} & $-0.85 \pm 0.15$
	& \hspace{0.1in} & $-1.96 \pm 1.32$
	& \hspace{0.1in} & $-3.53 \pm 5.06$
	& \hspace{0.1in} & $-2.50 \pm 1.01$
	& \hspace{0.1in} & $-1 \pm 1$			\\
\hline
\hspace{0.0in} & $d$ (AJM)
	& \hspace{0.1in}   $-1.0 \pm 0.1$
	& \hspace{0.1in} & $-0.92 \pm 0.27$
	& \hspace{0.1in} & $-0.85 \pm 0.14$
	& \hspace{0.1in} & $-0.81 \pm 0.64$
	& \hspace{0.1in} & $-0.52 \pm 0.78$
	& \hspace{0.1in} & $-0.49 \pm 0.14$
	& \hspace{0.1in} & $-1 \pm 1$			\\
\hline
\end{tabular}}
\label{tab:yr}
\end{center}
\end{table}
\end{savenotes}

Note that in Table~\ref{tab:yr} and in our numerical calculations we
make use of the latest values of the helicity amplitudes from the PDG
\cite{Beringer:2012zz}.  However, when comparing directly with the
GHRM analysis \cite{Gorchtein:2011mz} we will refer to the earlier,
2010 PDG values \cite{Nakamura:2010zzi} that were utilized by GHRM
for the $D_{13}(1520)$ and $P_{11}(1440)$ resonances.  The $y_R$
ratios using these earlier values are listed in parentheses in
Table~\ref{tab:yr}, but with errors evaluated using Gaussian
distributions.

For the nonresonant background, the models describing the
electromagnetic structure functions are transformed to the
$\gZ$ case according to the kinematic region considered.
For the region of low $Q^2$ but high $W$, the cross section in
the VMD+Regge model \cite{Alwall:2004wk} is modified using the
ratio in Eq.~(\ref{eq:GHRM45}), in analogy with Model~II of GHRM
\cite{Gorchtein:2011mz}.  However, instead of fixing the parameters
$\kappa_C^{T,L}$ so that the $\gg$ and $\gZ$ continuum pieces
are equal \cite{Gorchtein:2011mz}, we allow these to be determined
by demanding that the $\gZ$ structure functions be continuous
across the boundaries of this region, that is, at $W = 3$~GeV
and $Q^2 = 2.5$~GeV$^2$.
As we will see in the following section, this places strong
constraints on $\kappa_C^{T,L}$, leading to significantly
reduced uncertainties on the resulting value of \regzv.

Finally, the $\gZ$ structure functions in the DIS region, at
$W^2 > 4$~GeV$^2$ and $Q^2 > 2.5$~GeV$^2$, are computed from the
ABM11 PDF parametrization \cite{Alekhin:2012ig, Alekhin:priv}.
The transformation from $\gg$ to $\gZ$ is trivial at the parton
level, amounting to a replacement of the quark electric charges $e_q$
multiplying the universal PDFs by the weak vector charges $g_V^q$.
In the absence of $\gZ$ structure function data at low $Q^2$, the
relative magnitude of the higher twist corrections to $\ftgz$ was
taken \cite{Alekhin:priv} to be the same as for $\ftgg$.
To account for this uncertainty, we therefore assign a conservative
5\% uncertainty on $\fogz$ and $\ftgz$ over the entire range of
kinematics in Region~III.  Since it is given by a difference of
the $\ftgz$ and $\fogz$ structure functions (see Eq.~(\ref{eq:FL})),
the longitudinal structure function $\flgz$ will necessarily have
a larger relative uncertainty.

\section{Phenomenological constraints}
\label{sec:constraints}

As mentioned in the previous section, the central value of \regzv
in Ref.~\cite{Gorchtein:2011mz} is given by the average of Models~I
and II, with the dominant nonresonant background contribution taken 
from Model~II.
If it were possible to reduce the background uncertainty, the error
on the final \regzv correction could also be lowered significantly.

In their calculation of \bgZ, GHRM \cite{Gorchtein:2011mz} estimate
the $\gZ$ nonresonant background cross section by transforming the
$\gg$ cross section in the VMD+Regge model \cite{Alwall:2004wk}
according to
\bea
\sigma_{T,L}^{\gZ \rm(bgd)}
&=& \left( \frac{\sigma_{T,L}^{\gZ}}{\sigma_{T,L}^{\gg}} \right)\,
    \sigma_{T,L}^{\rm VMD},
\eea
with the electromagnetic cross sections $\sigma_{T,L}^{\rm VMD}$
parametrized as in Eqs.~(\ref{eq:sTv}) and (\ref{eq:sLv}), and
the rescaling factor $(\sigma_{T,L}^{\gZ}/\sigma_{T,L}^{\gg})$
given by Eq.~(\ref{eq:GHRM45}).
The uncertainties on the $\gZ$ cross section are obtained by comparing
each $R_V^{T, L}$ ratio in Eq.~(\ref{eq:GHRM45}) with HERA data on
exclusive vector meson electroproduction \cite{Breitweg:2000mu}
({\it cf.} Fig.~13 of Ref.~\cite{Gorchtein:2011mz}), with the
uncertainty taken to be the difference between the two.

\begin{figure}[t]
\includegraphics[width=\textwidth]{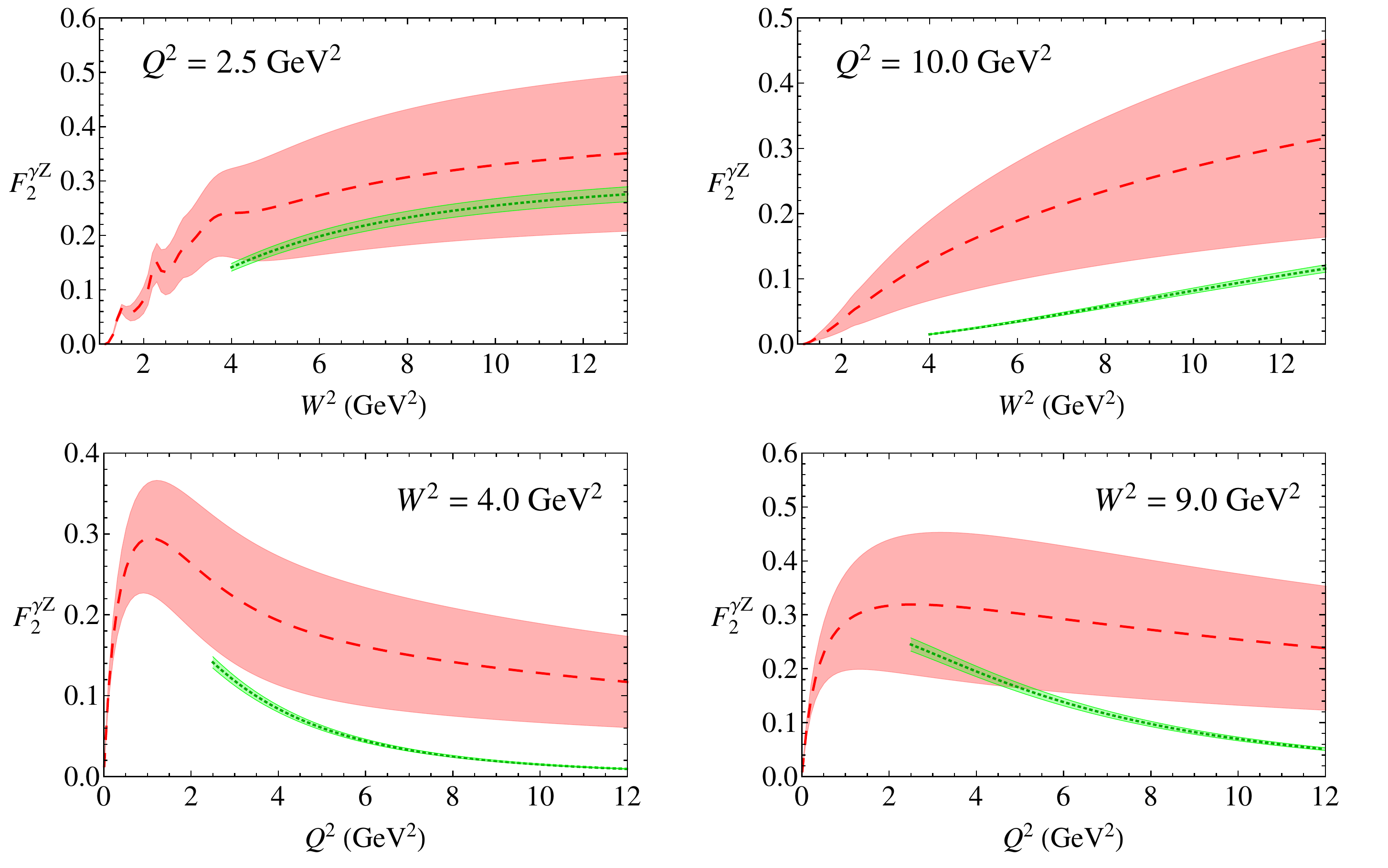}
\caption{(color online)
	Comparison of the proton $\ftgz$ structure function in the
	VMD+Regge model (Model~II) of GHRM \cite{Gorchtein:2011mz}
	(red dashed) with the ABM11 global parametrization
	\cite{Alekhin:2012ig} (green dotted), for fixed $Q^2$
	(top panels) and fixed $W^2$ (bottom panels).
	Note that the VMD+Regge model only includes uncertainties
	from the continuum part of the background, while the
	ABM11 parametrization includes an overall 5\% error.}
\label{fig:F2gZ}
\end{figure}

The final contribution to the background error comes from the values
of $\kappa_C^{T, L}$ in Eq.~(\ref{eq:GHRM45}).  In the GHRM analysis
\cite{Gorchtein:2011mz} this term is equated with the electromagnetic
continuum piece, assuming a 100\% uncertainty.  The resulting $\ftgz$
structure function is illustrated in Fig.~\ref{fig:F2gZ} as a function
of both $W^2$ and $Q^2$, and compared with the ABM11 global fit
\cite{Alekhin:2012ig}.  Note that the uncertainty band on the GHRM
VMD+Regge calculation includes only the continuum part of the
background, and will be larger once the resonant uncertainty is
included.  The comparison clearly shows that the GHRM uncertainties
are significantly larger than those typically obtained from global
QCD analyses, especially in the region of intermediate $W$ and $Q^2$
where both descriptions should be valid.
Furthermore, as suggested already in Figs.~\ref{fig:F2gCP} and
\ref{fig:F2gVP}, the central values lie systematically above the
PDF parametrizations.

Although the VMD model itself does not provide any additional
constraints on the interference continuum contribution, we shall
examine in this section the possibility of constraining
$\kappa_C^{T, L}$ using existing knowledge of parton distributions,
as well as recent data on parity-violating inelastic scattering
from the Jefferson Lab E08-011 experiment \cite{E08011-RES}.
These constraints will make it possible to reduce the overall
uncertainty in \regzv compared with those obtained in earlier
analyses, Eq.~(\ref{eq:ReBox_other}).

\subsection{Constraints from PDFs}
\label{ssec:PDFs}

In the deep-inelastic region at high $W$ ($W \gtrsim 2$~GeV) and $Q^2$
($Q^2 \gtrsim 1$~GeV$^2$), structure functions can be described in terms
of leading twist PDFs, with corrections from target mass and higher
twist contributions included to account for residual, $1/Q^2$-suppressed
nonperturbative effects.  While a PDF-based description will eventually 
break down at low $W$ and $Q^2$, the region where the continuum
contributions to the cross sections are relevant overlaps with the
typical reach of global PDF parametrizations \cite{Accardi:2011fa,
Owens:2012bv, Alekhin:2012ig, Ball:2010de, JimenezDelgado:2009tv,
Martin:2009iq}.  One can therefore constrain the nonresonant part of
the $\gZ$ structure functions by requiring consistency of the model
in the overlap region with the PDF parametrizations.

Our fit of the parameters $\kappa_C^{T,L}$ involves equating the
cross section ratios $\sigma_{T,L}^{\gZ}/\sigma_{T,L}^{\gg}$ in
Eq.~(\ref{eq:GHRM45}) with the structure function ratios computed
from global QCD fits in the DIS region
[see Eqs.~(\ref{eq:sig_def}) and (\ref{eq:FL})],
\bea 
\frac{\sigma_T^{\gZ}}{\sigma_T^{\gg}}
&=& \left. \frac{\fogz}{\fogg} \right|_{\rm DIS},
\hspace*{1.5cm}
\frac{\sigma_L^{\gZ}}{\sigma_L^{\gg}}\
 =\ \left. \frac{\flgz}{\flgg} \right|_{\rm DIS},
\label{eq:sigTFi}
\eea
where the DIS structure functions $F_{1,L}^{\gg,\gZ}$ are taken from
the ABM11 parametrization \cite{Alekhin:2012ig}.  As discussed in
Sec.~\ref{sec:SFs}, in fitting $\kappa_C^{T,L}$ in the DIS region,
to be conservative we assume an overall 5\% uncertainty on $\fogz$,
and a 40\% uncertainty on $\flgz$, which exceeds the uncertainties
quoted in Ref.~\cite{Alekhin:2012ig} over the kinematics relevant
for the \regzv calculation.

\begin{figure}[t]
\includegraphics[width=3.19in]{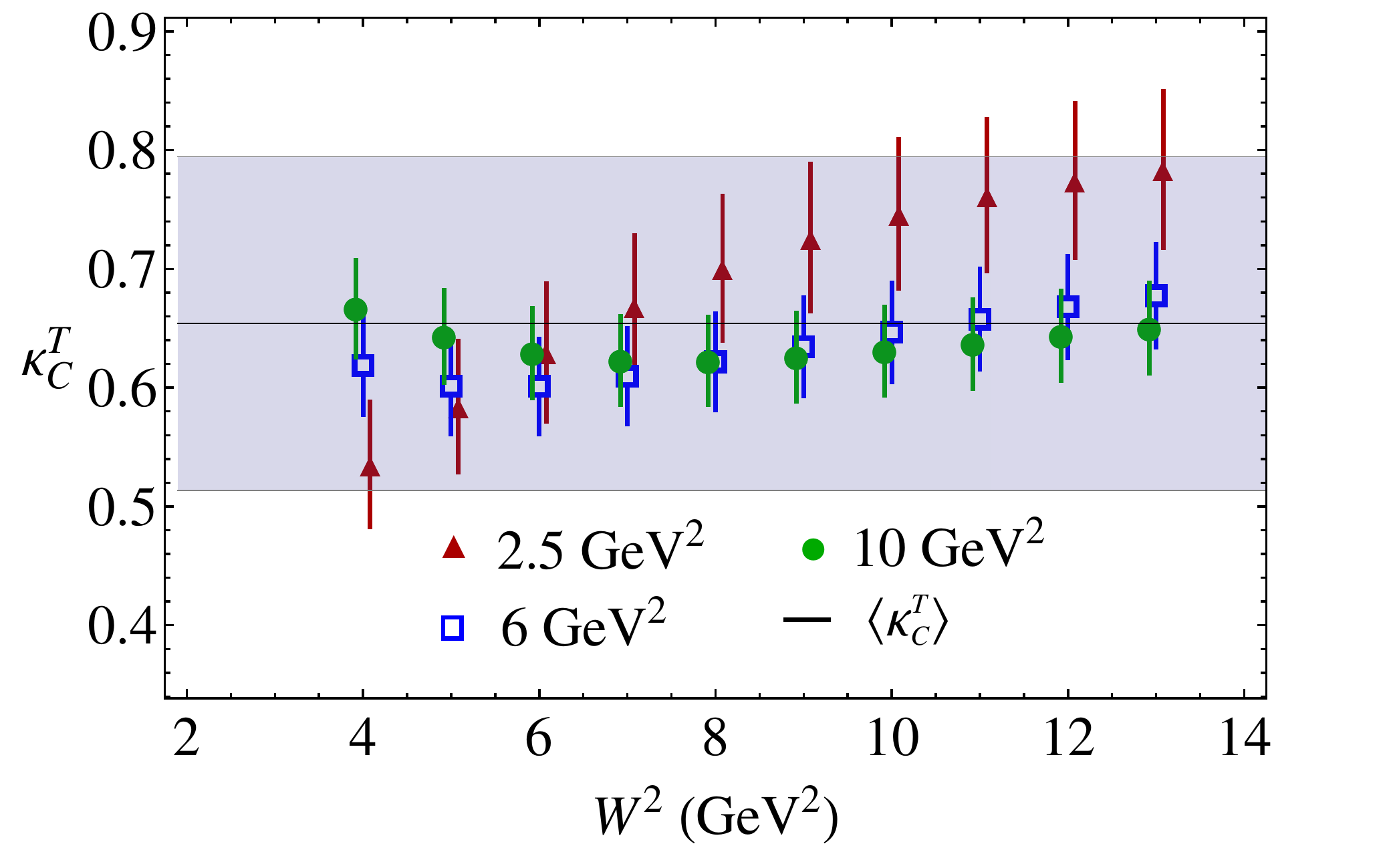}
\includegraphics[width=3.23in]{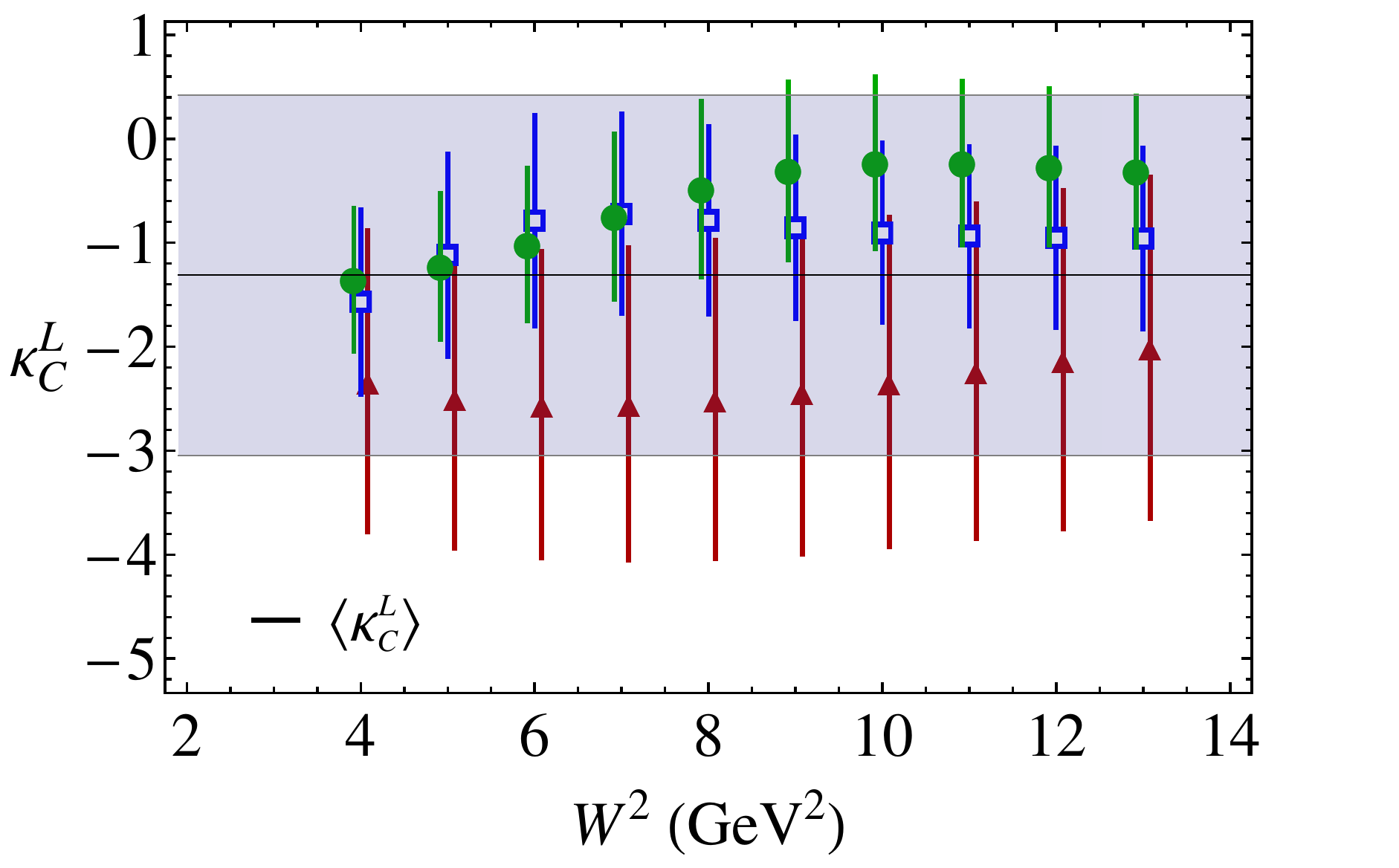}
\caption{(color online)
	Continuum parameters $\kappa_C^T$ (left) and $\kappa_C^L$
	(right) fitted to the DIS data, parametrized by the ABM11
	global QCD fit \cite{Alekhin:2012ig}, as a function of
	$W^2$ for fixed $Q^2 = 2.5$~GeV$^2$ (red triangles),
	6~GeV$^2$ (blue squares), and 10~GeV$^2$ (green circles).
	The average values $\langle \kappa_C^{T,L} \rangle$ are
	indicated by the solid lines, with the shaded band giving
	their uncertainty.  Note that some of the points have been
	slightly offset for clarity.}
\label{fig:chi2}
\end{figure}

For the constrained fit we determine the values of $\kappa_C^{T,L}$
that minimize the $\chi^2$ for each point in $W^2$ and $Q^2$, over
a range of $W^2$ values at fixed $Q^2$ near the boundary between the
DIS region (Region~III) and the other regions in Fig.~\ref{fig:W2Q2}.
To test the stability of the fitted $\kappa_C^{T,L}$ values with
respect to the matching scale, we consider several different values
of $Q^2$ ($Q^2 = 2.5$, 6 and 10~GeV$^2$).  The resulting fits in
Fig.~\ref{fig:chi2} indicate relatively mild dependence on the scale,
which becomes negligible with increasing $Q^2$ for $\kappa_C^T$,
but with the expected larger uncertainties for $\kappa_C^L$.

The central values of $\kappa_C^{T,L}$ are computed by averaging over
the three sets of $Q^2$ values, and the uncertainty determined by taking
into account both the $W^2$ dependence of the fits and the PDF error.
Because the $\kappa_C^{T,L}$ values at the different $Q^2$ are
correlated, performing a simple $\chi^2$ fit to all the sets may
underestimate the errors.  As a more reliable error estimate, we
combine in quadrature the uncertainties arising from
  (i) the $W^2$ dependence, for which we take the average of the
difference between the central values of the lowest and highest
points for the $Q^2$ set giving the strongest $W^2$ dependence
(namely, for $Q^2 = 2.5$~GeV$^2$ for $\kappa_C^T$, and
	     $Q^2 = 10$~GeV$^2$  for $\kappa_C^L$); and
  (ii) the PDF error, the uncertainty for which is given by the
data point with the largest error in the entire set (which occurs
for $Q^2 = 2.5$~GeV$^2$ for both $\kappa_C^T$ and $\kappa_C^L$).
The final fitted values of the continuum parameters are found to be
\be 
\kappa_C^T =  0.65 \pm 0.14\, , \qquad \qquad
\kappa_C^L = -1.3 \pm 1.7\, .
\label{eq:kappaC}
\ee
Compared with the uncertainties assumed by GHRM \cite{Gorchtein:2011mz}
our uncertainty on the transverse parameter $\kappa_C^T$ is about five
times smaller, while that on the longitudinal parameter $\kappa_C^L$
is almost two and a half times larger.  However, the error on
$\kappa_C^L$ has minimal effect on the $\gZ$ cross section at these
kinematics because of the relatively small contribution of the
longitudinal structure function.

\begin{figure}
\includegraphics[width=\textwidth]{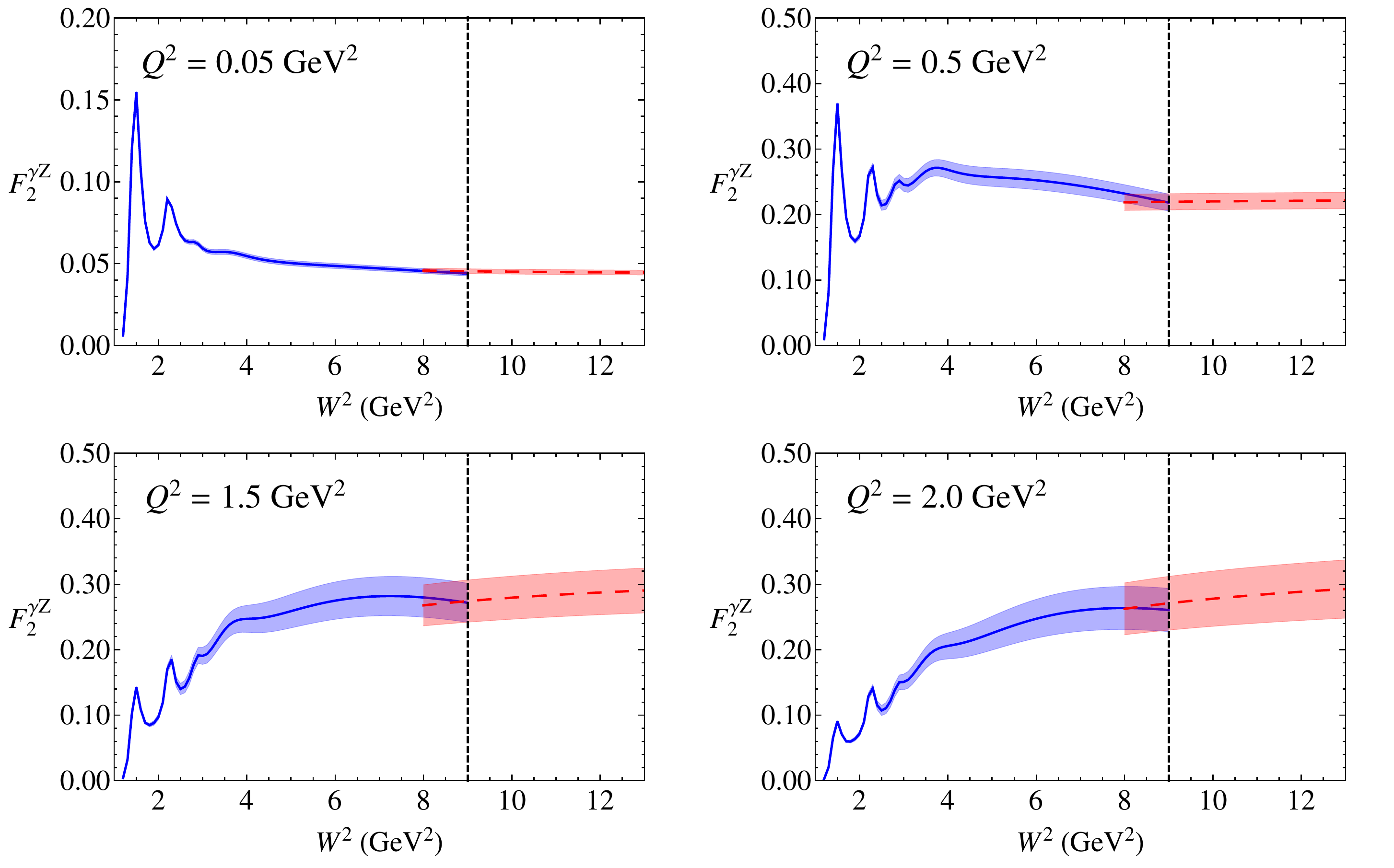}
\includegraphics[width=\textwidth]{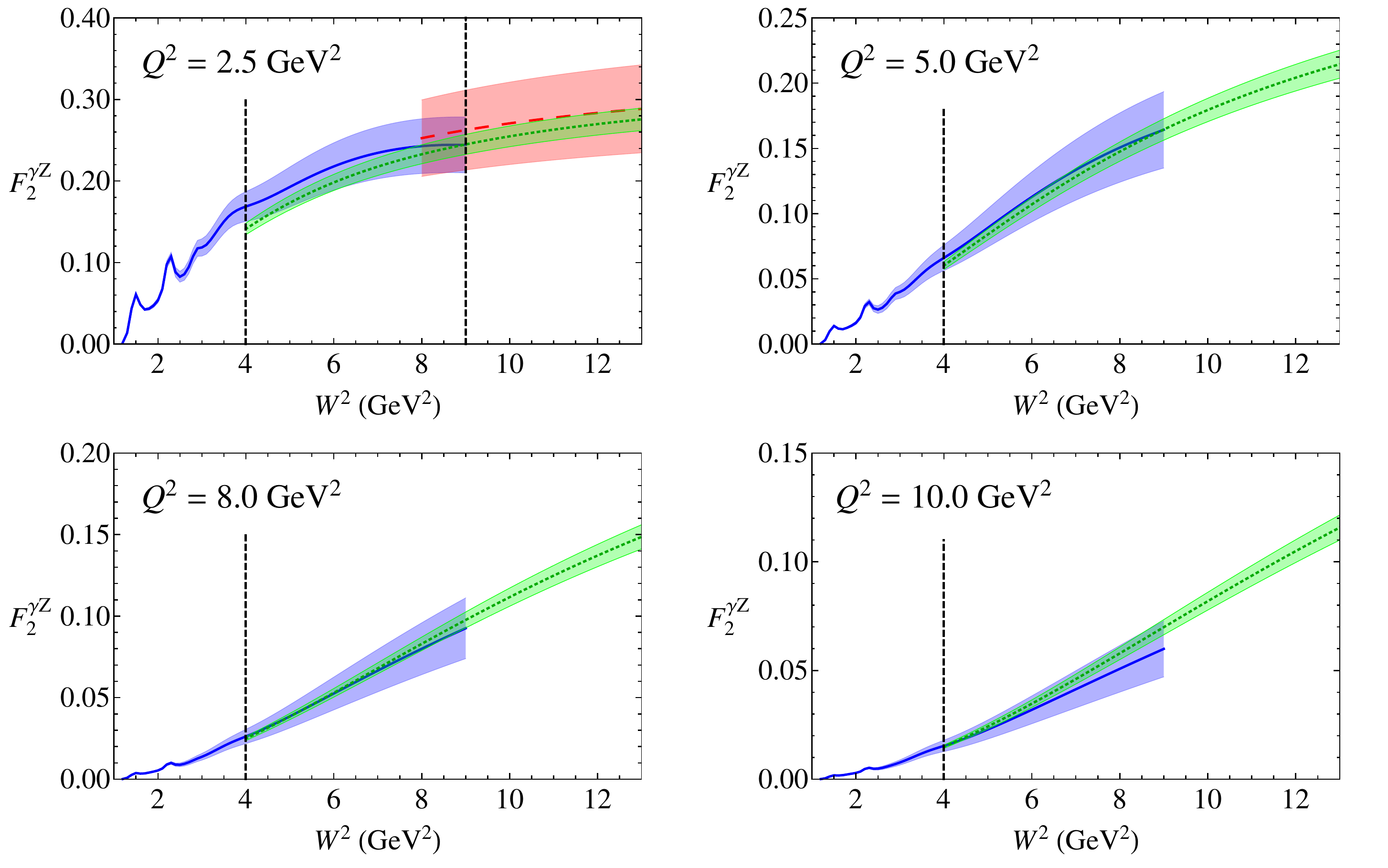}
\caption{(color online)
	Proton $\ftgz$ structure function versus $W^2$ at various
	fixed $Q^2$ values for the low-$W$ CB fit \cite{Christy:2007ve}
	(blue solid) and the high-$W$ VMD+Regge \cite{Alwall:2004wk}
	(red dashed) and ABM11 \cite{Alekhin:2012ig} (green dotted)
	parametrizations.  The boundaries between the Regions~I, II
	and III are indicated by the vertical lines at $W^2=4$ and
	9~GeV$^2$.}
\label{fig:gZ1}
\end{figure}

\begin{figure} 
\includegraphics[width=\textwidth]{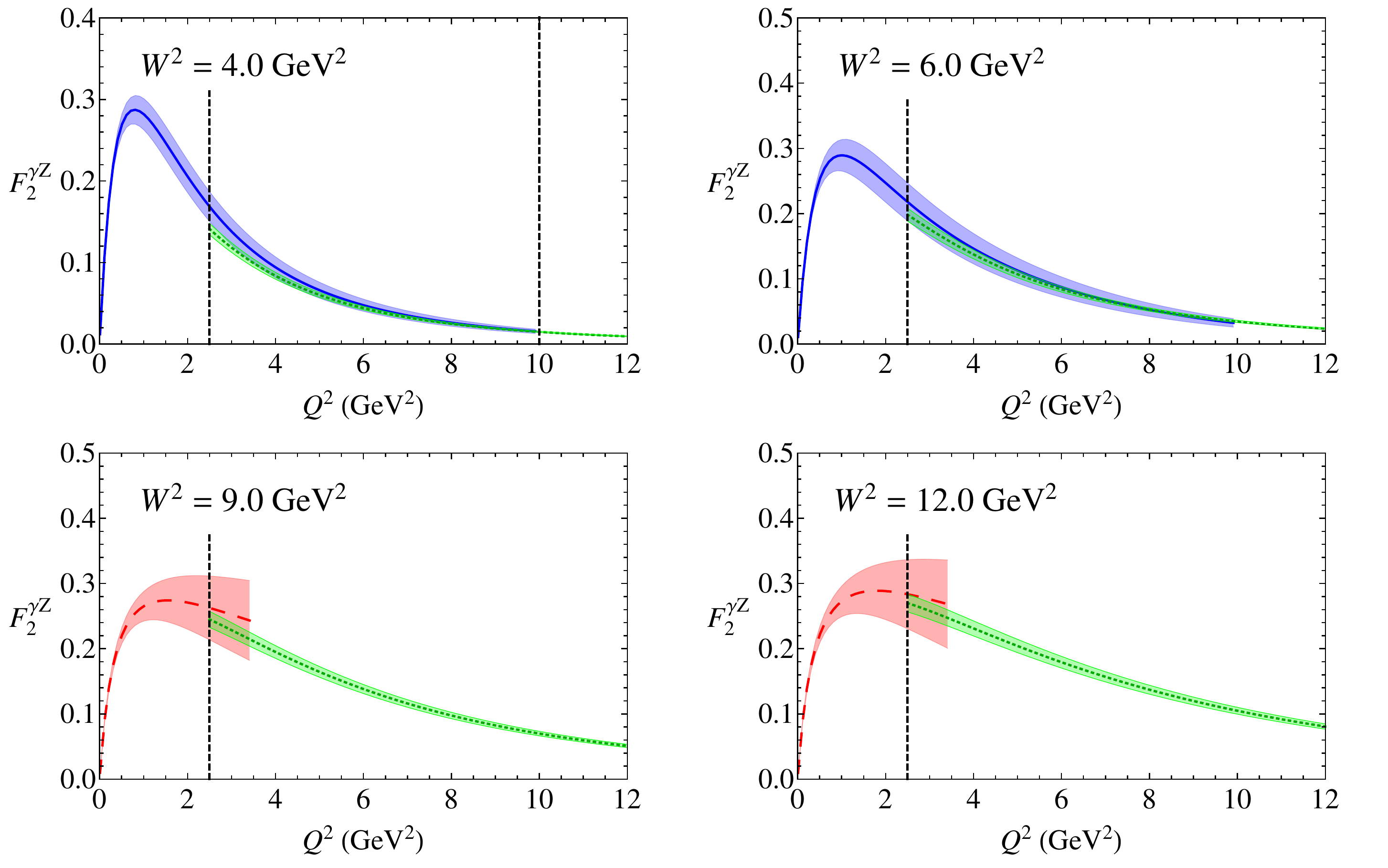}
\caption{(color online)
	Proton $\ftgz$ structure function versus $Q^2$ at
	fixed $W^2=4$, 6, 9 and 12~GeV$^2$ for the CB fit
	\cite{Christy:2007ve} (blue solid), the ABM11 PDF
	parametrization \cite{Alekhin:2012ig} (green dotted),
	and the VMD+Regge model \cite{Alwall:2004wk} (red dashed),
	with the boundaries between Regions~I, II and III
	indicated by the vertical lines at fixed $Q^2$.}
\label{fig:gZ2}
\end{figure}

The resulting $\ftgz$ structure function with the constrained
$\kappa_C^{T,L}$ values is shown in Fig.~\ref{fig:gZ1} for
fixed $Q^2$, ranging from $Q^2 = 0.05$ to 10~GeV$^2$.
The models of the $\gZ$ structure functions are seen to match
very well at the boundaries between the Regions~I, II and III.
As for the interference $\ftgz$ structure function in
Fig.~\ref{fig:F2gZ}, only the continuum
uncertainty is included in these examples; 
this allows a direct comparison with the uncertainty in the GHRM
model input which dominates all other
uncertainties.  The comparison between Figs.~\ref{fig:F2gZ} and
\ref{fig:gZ1} at the corresponding kinematics illustrates the
significant reduction in the $\ftgz$ uncertainty that results from
constraining the structure functions by the global QCD fits of PDFs.
A similarly large reduction in the uncertainty can be seen in
Fig.~\ref{fig:gZ2} for $\ftgz$ as a function of $Q^2$ at fixed
$W^2$ values.

The remaining uncertainty on the background contribution is
associated with the $R_\omega^{T, L}$ and $R_\phi^{T, L}$ terms
in Eq.~(\ref{eq:GHRM45}).  Following GHRM \cite{Gorchtein:2011mz},
we take the difference between these ratios calculated in the
VMD+Regge model at $Q^2 = 7$~GeV$^2$ and the experimental vector
meson cross sections from HERA \cite{Breitweg:2000mu}, assuming
$R_\omega^T = R_\omega^L$ and $R_\phi^T = R_\phi^L$ (see Fig.~13
of \cite{Gorchtein:2011mz}).
This uncertainty is then added in quadrature with the continuum
uncertainty, along with the resonance contribution discussed in
Sec.~\ref{sec:SFs}, to obtain the total error on the $\gZ$
structure functions used in estimating \regzv.

\begin{figure}[ht]
\vspace*{0.2cm}
\includegraphics[width=3.2in]{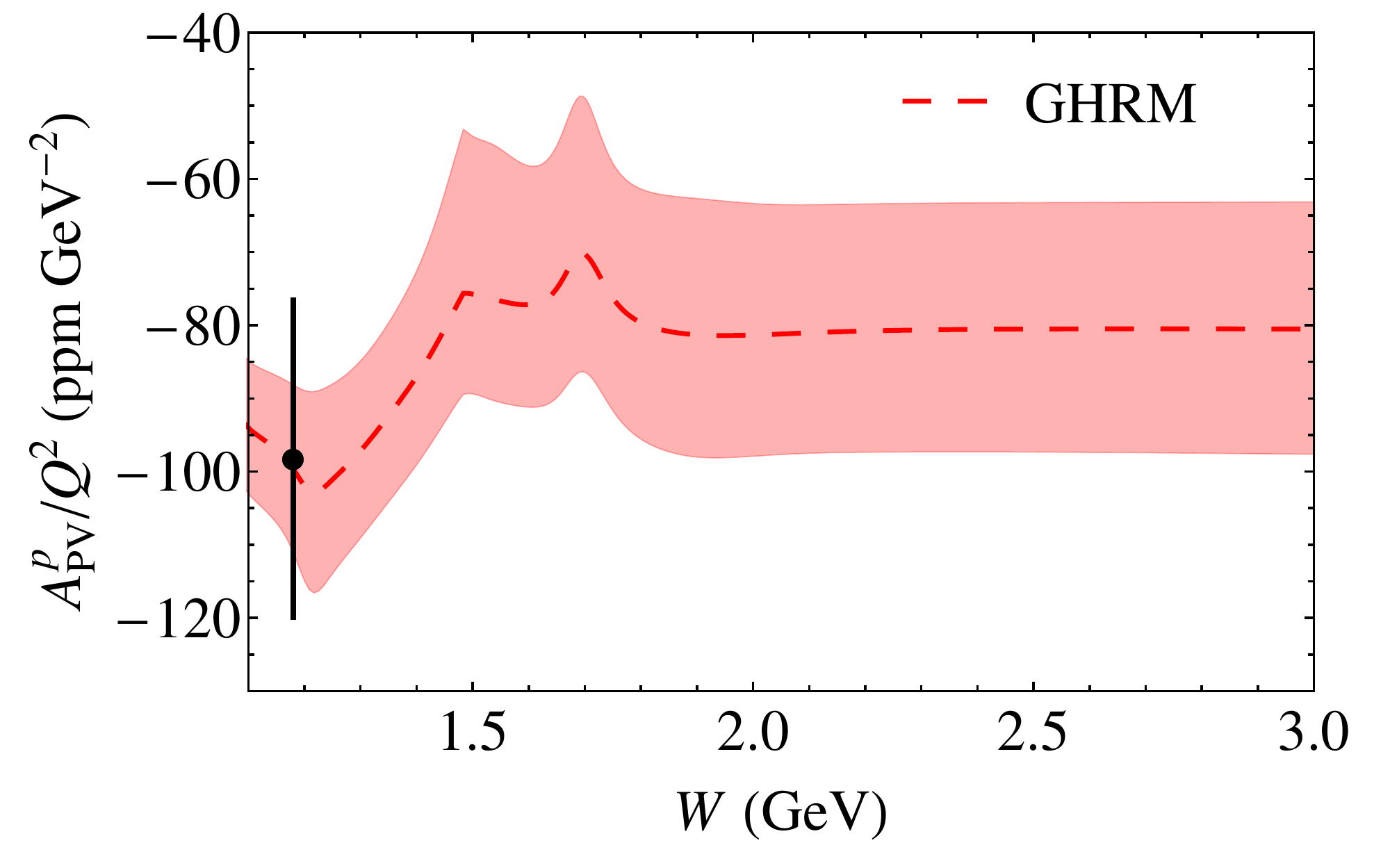}
\includegraphics[width=3.2in]{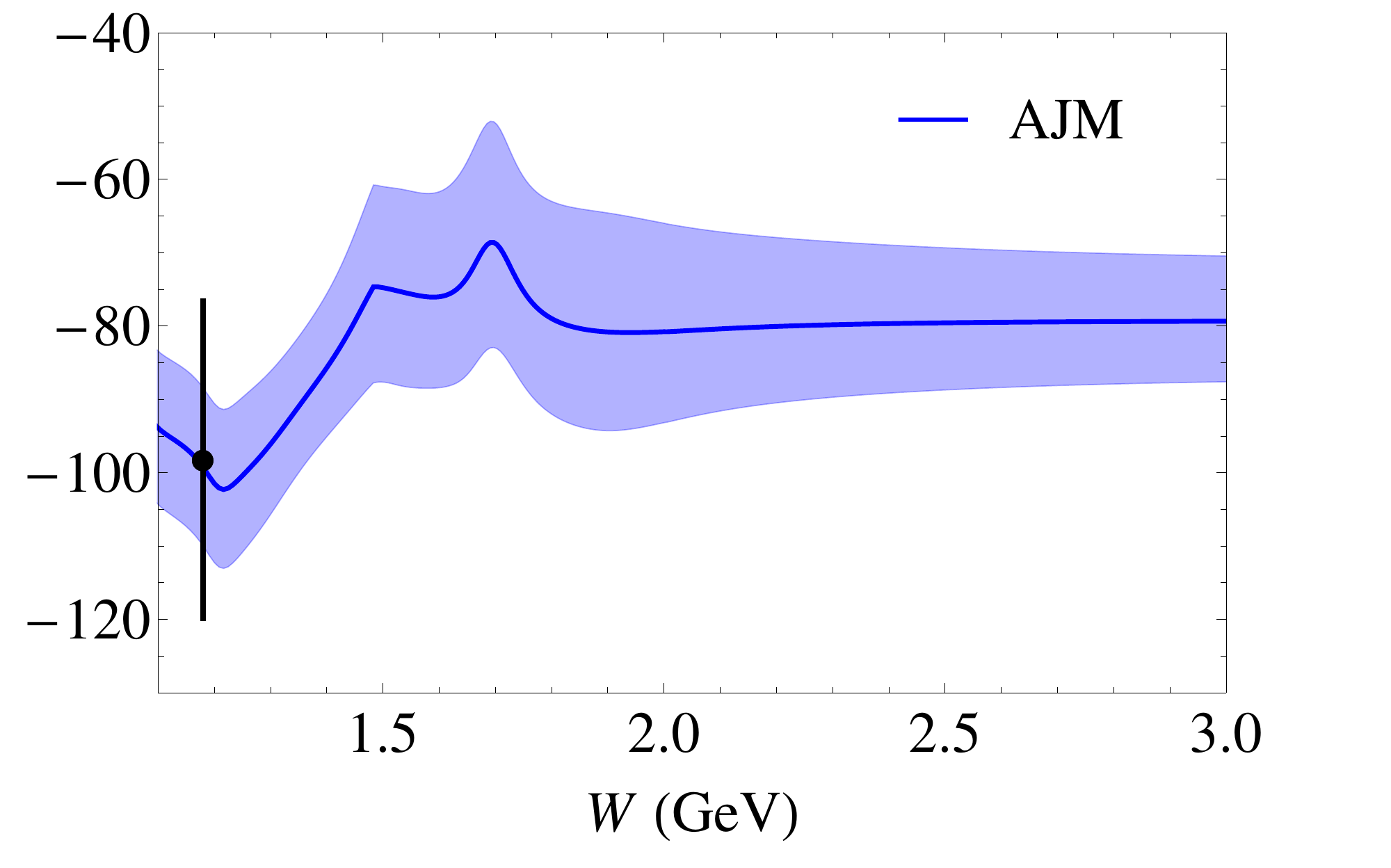}%
\caption{(color online)
	Proton parity-violating inelastic asymmetry $A_{\rm PV}/Q^2$
	as a function of $W$, at fixed incident energy $E = 0.69$~GeV
	and $Q^2 = 0.34$~GeV$^2$, for the GHRM Model~II
	\cite{Gorchtein:2011mz} (left) and the AJM model (right).
	The data point at $W = 1.18$~GeV (black circle) is from
	the Jefferson Lab G0 experiment \cite{Androic:2012doa}.\\}
\label{fig:G0apvProMII}
\includegraphics[width=3.2in]{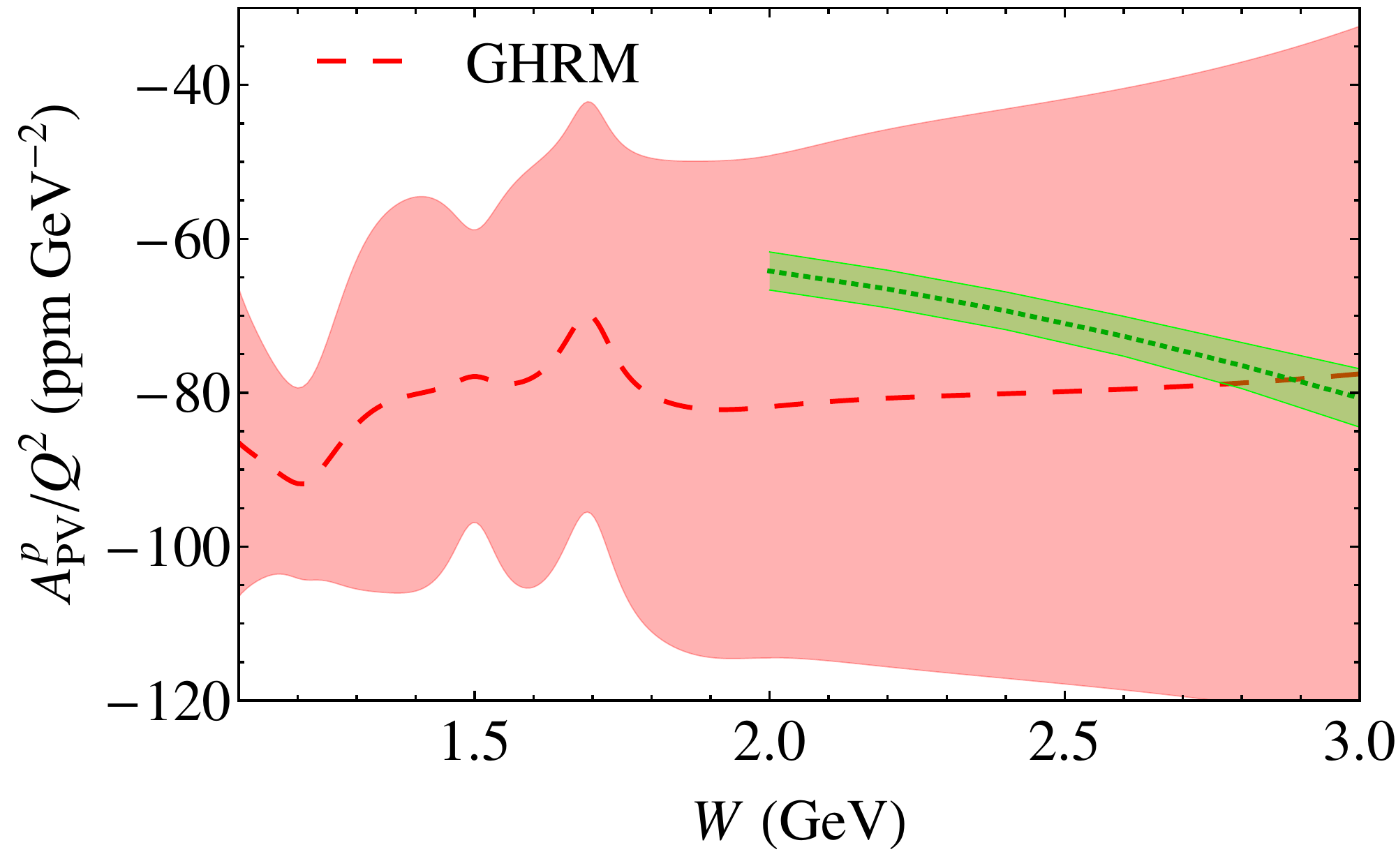}
\includegraphics[width=3.2in]{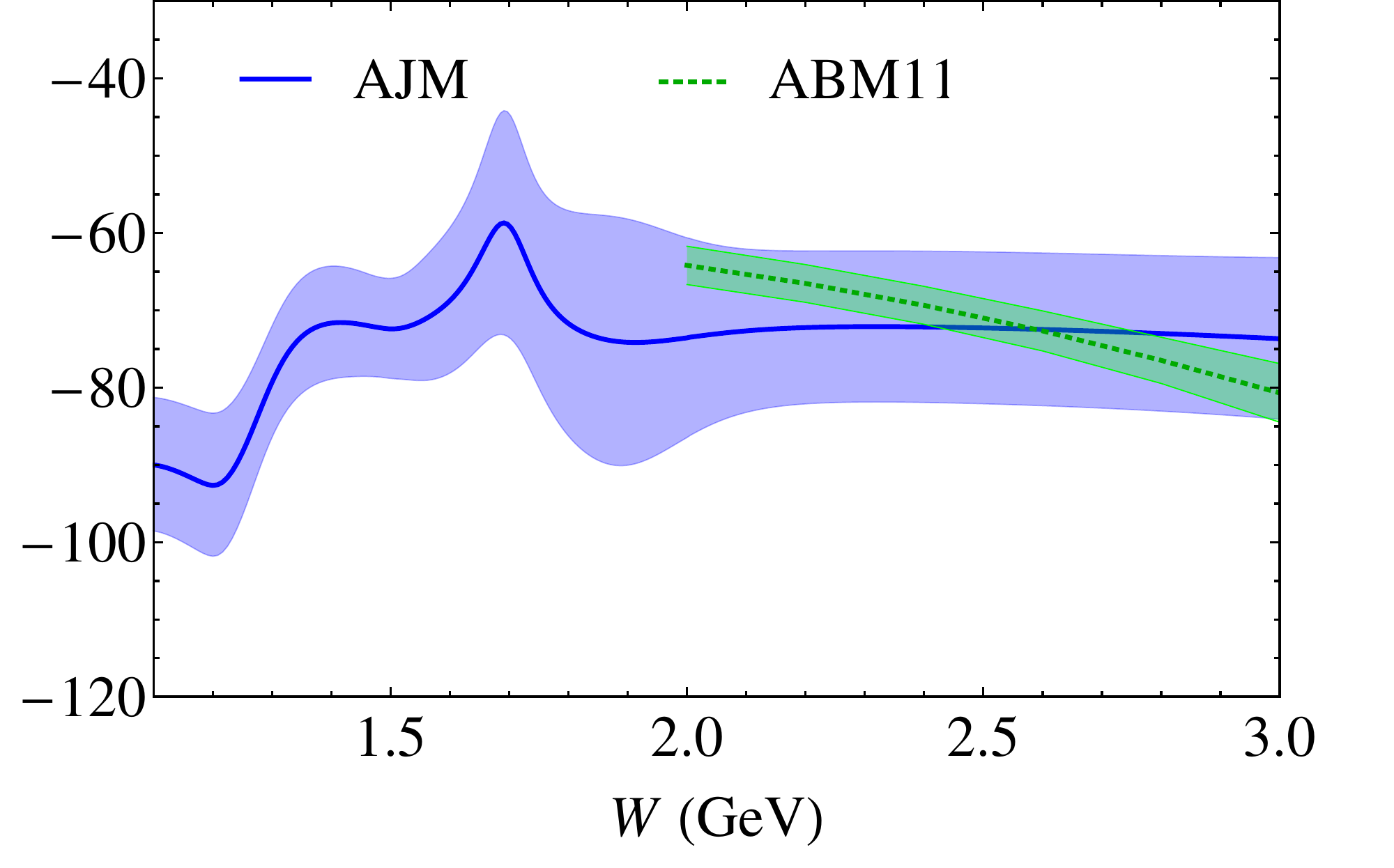}%
\caption{(color online)
	Proton parity-violating inelastic asymmetry $A_{\rm PV}/Q^2$
	as a function of $W$, at fixed incident energy $E = 6$~GeV
	and $Q^2 = 2.5$~GeV$^2$, for the GHRM Model~II
	\cite{Gorchtein:2011mz} (left) and the AJM model (right).
	The asymmetry computed directly from PDFs \cite{Alekhin:2012ig}
	is represented by the green band.}
\label{fig:apvProMII}
\end{figure}

The impact of the total uncertainty reduction is illustrated
in Figs.~\ref{fig:G0apvProMII} and \ref{fig:apvProMII} for the
parity-violating inelastic asymmetry for the proton,
\bea
A_{\rm PV}
&=& g_A^e {\left( \frac{G_F Q^2}{2 \sqrt{2} \pi \alpha} \right)
\frac{x y^2 \fogz
      + \left( 1 - y - \displaystyle{x^2 y^2 M^2 \over Q^2} \right)
        \ftgz
      + \displaystyle{g_V^e \over g_A^e}
        \left( y - \frac12 y^2 \right) x \fthgz}
     {x y^2 \fogg
      + \left( 1 - y - \displaystyle{x^2 y^2 M^2 \over Q^2} \right)
        \ftgg}},
\label{eq:apvdisd}
\eea
where $y = \nu/E$ is the fractional energy transferred to the target.
In addition to the vector $F_{1,2}^{\gZ}$ structure functions, the
asymmetry $A_{\rm PV}$ depends also on the axial-vector
$\fthgz$ structure function.  For the the resonance contribution to
$\fthgz$ we use the parametrization of the axial-vector transition
form factors of Lalakulich {\it et al.} \cite{Lalakulich:2005cs,
Lalakulich:2006sw, Lalakulich:2006yn}.  For the background we follow
Ref.~\cite{Carlson:2012yi} and rescale the electromagnetic cross
sections \cite{Christy:2007ve} by the average of the $x \to 0$
and SU(6) quark model limits, which gives $\fthgz = 5/3\, \fogg$.
(Note that for the deuteron this average becomes $\fthgz = 9/5\, \fogg$.)

The asymmetries calculated in the AJM and GHRM models are shown in
Fig.~\ref{fig:G0apvProMII} at an incident energy $E = 0.69$~GeV and
$Q^2 = 0.34$~GeV$^2$, corresponding to the kinematics of the recent
G0 measurement at Jefferson Lab near the $\Delta$ resonance region
\cite{Androic:2012doa}.  The central values of both models agree well
with the data, although the experimental uncertainty is too large to
enable meaningful constraints to be placed on the $\gZ$ structure
functions.  The constraint on the $\kappa_C^T$ value from matching to
the DIS structure functions in the AJM model renders the uncertainty
band somewhat smaller than the GHRM uncertainty \cite{Gorchtein:2011mz}
at higher values of $W$.
(Note that the uncertainty on $A_{\rm PV}$ is computed
by taking the upper and lower values of the input $\gZ$ structure
functions, and is therefore asymmetric.)

The difference in the error bands becomes more pronounced at larger
$Q^2$, as seen in Fig.~\ref{fig:apvProMII} at $E = 6$~GeV and
$Q^2 = 2.5$~GeV$^2$, which are representative of typical kinematics
at Jefferson Lab (see Sec.~\ref{ssec:Ad} below).
Here the uncertainty on the GHRM model asymmetry at $W \sim 2$~GeV
is around four times larger than the corresponding uncertainty on the
constrained AJM model asymmetry.
For comparison, we also show in Fig.~\ref{fig:apvProMII} the asymmetry
computed directly from PDFs \cite{Alekhin:2012ig} in the region
$W > 2$~GeV where a partonic description is expected to be valid.

The uncertainty in the PDF-based calculation is slightly smaller than,
but qualitatively similar to, that in the AJM model, while the GHRM
model uncertainty is significantly overestimated in the region of
overlap.  We stress that although the DIS region makes only a modest
contribution to \regzv, the requirement that the $\gZ$ cross sections
match across the DIS-resonance region boundary imposes strong
constraints on the $\gZ$ structure functions also at lower $W$ and
$Q^2$.  In the following section we confront this against new data on
parity-violating electron-deuteron scattering in the resonance region.

\subsection{Deuteron asymmetry}
\label{ssec:Ad}

The E08-011 experiment \cite{Xiaochao:priv, E08011-RES} at Jefferson Lab recently measured
the parity-violating asymmetry in inclusive electron-deuteron scattering
over a range of $W$ and $Q^2$ in both the resonance and DIS regions.
While the DIS region data are currently still being analyzed
\cite{Xiaochao:priv}, the available resonance region data
\cite{E08011-RES} can be used to provide an independent test of
the procedure for estimating the $\gZ$ structure functions.
This is particularly important for \regzv, since the integrals in
Eq.~(\ref{eq:ImBoxV}) are dominated by Region~I in Fig.~\ref{fig:W2Q2}.

\begin{figure}[ht]
\includegraphics[width=3.2in]{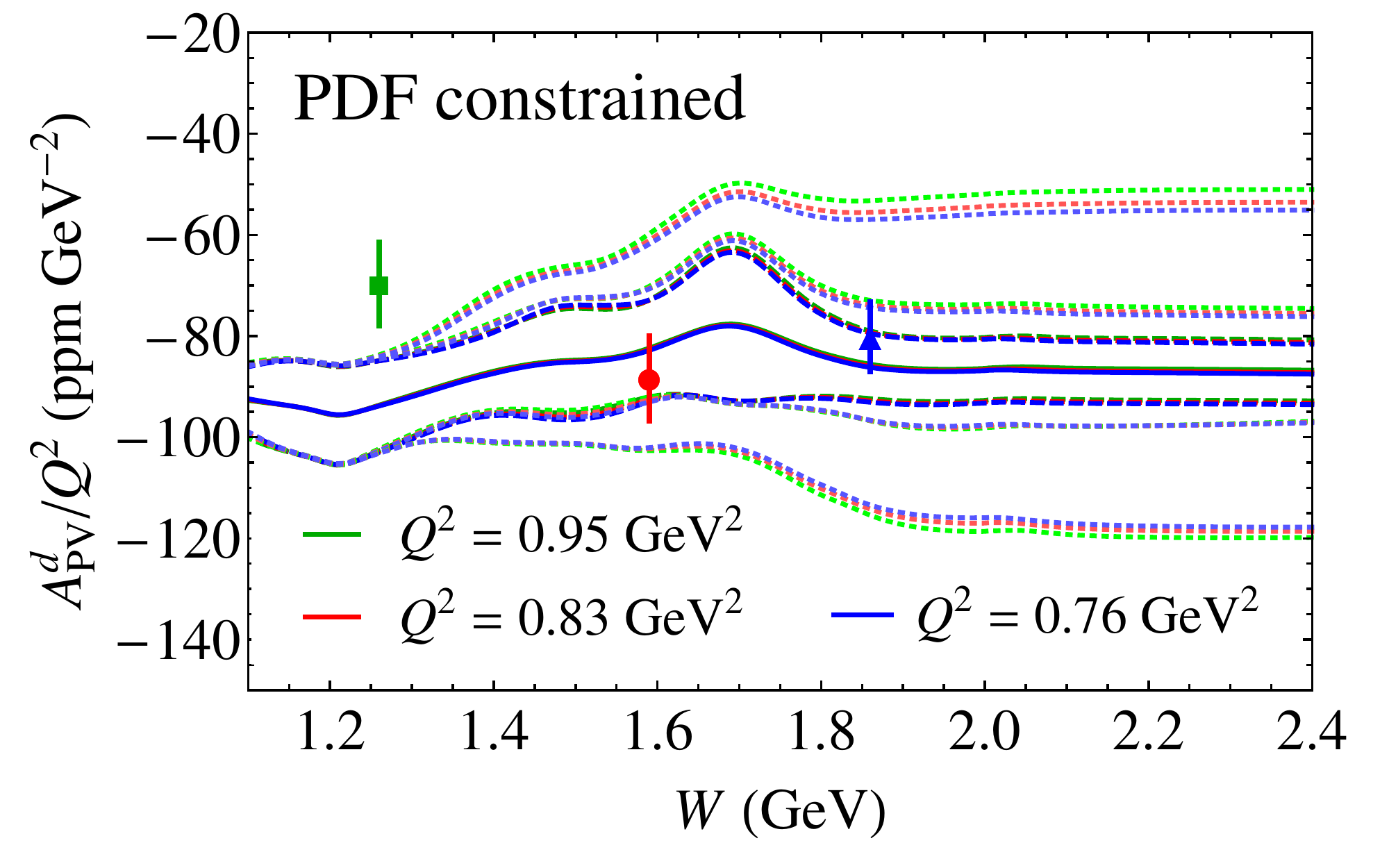}%
\includegraphics[width=3.2in]{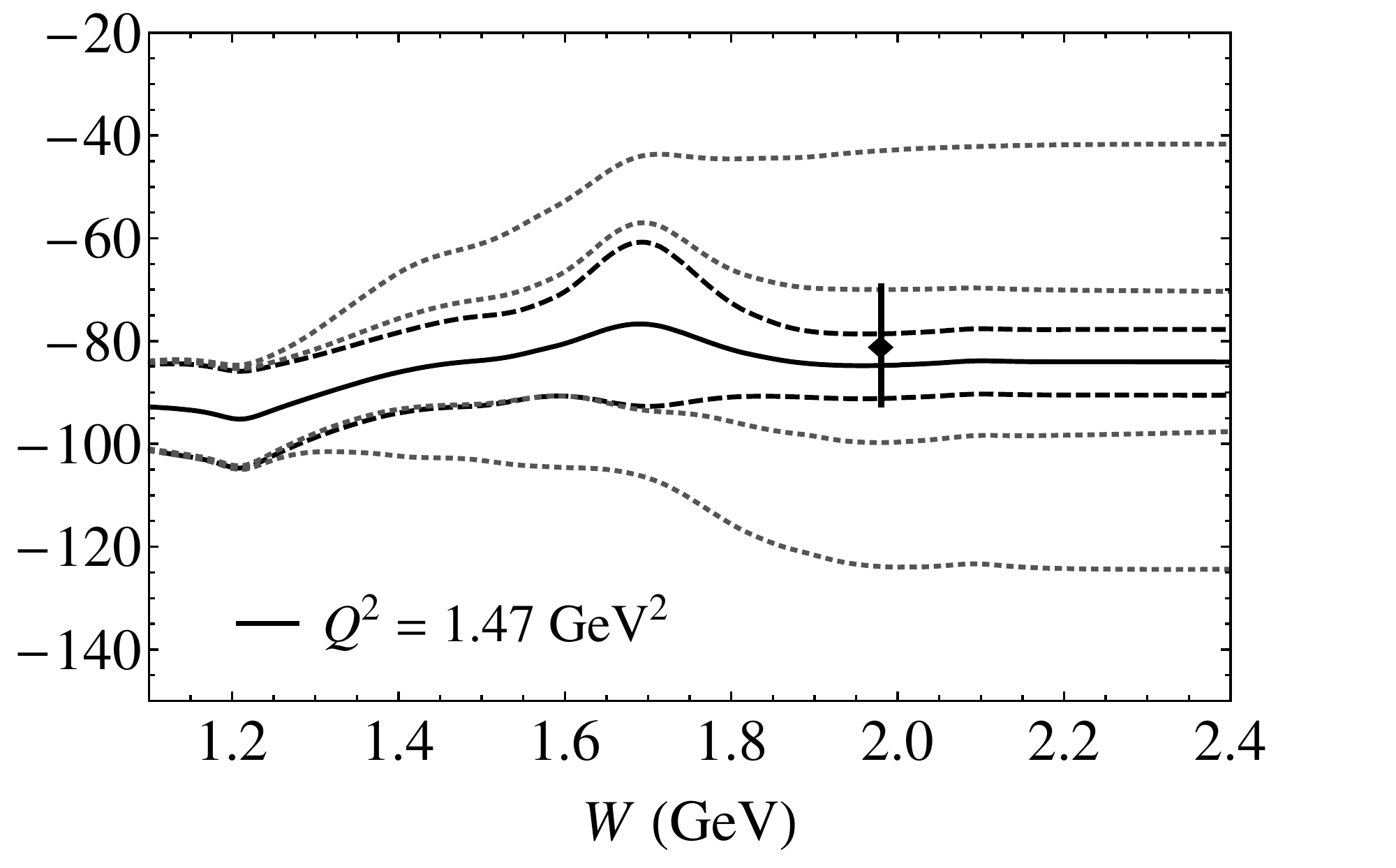}
\caption{(color online)
	Deuteron parity-violating asymmetry $A_{\rm PV}^d/Q^2$
	as a function of $W$ for incident electron energy
	$E = 4.9$~GeV (left) and $E = 6.1$~GeV (right).
	The data points from the Jefferson Lab E08-011 experiment
	\cite{E08011-RES} at $W = 1.26$ (green square), 1.59 (red circle),
	1.86 (blue triangle) and 1.98~GeV (black diamond) correspond
	to average values of $Q^2 = 0.95$, 0.83, 0.76 and 1.47~GeV$^2$,
	respectively.  The AJM model uncertainties (inner dashed band)
	are constrained by matching the continuum parameters
	$\kappa_C^{T,L}(d)$ to the DIS region $\gZ$ structure functions
	\cite{Alekhin:2012ig}, and are compared with those computed with
	errors on $\kappa_C^{T,L}(d)$ of 100\% (outer dotted bands) and
	25\% (inner dotted bands).\\}
\label{fig:apvcp}
\end{figure}

\begin{figure}[h]
\includegraphics[width=3.2in]{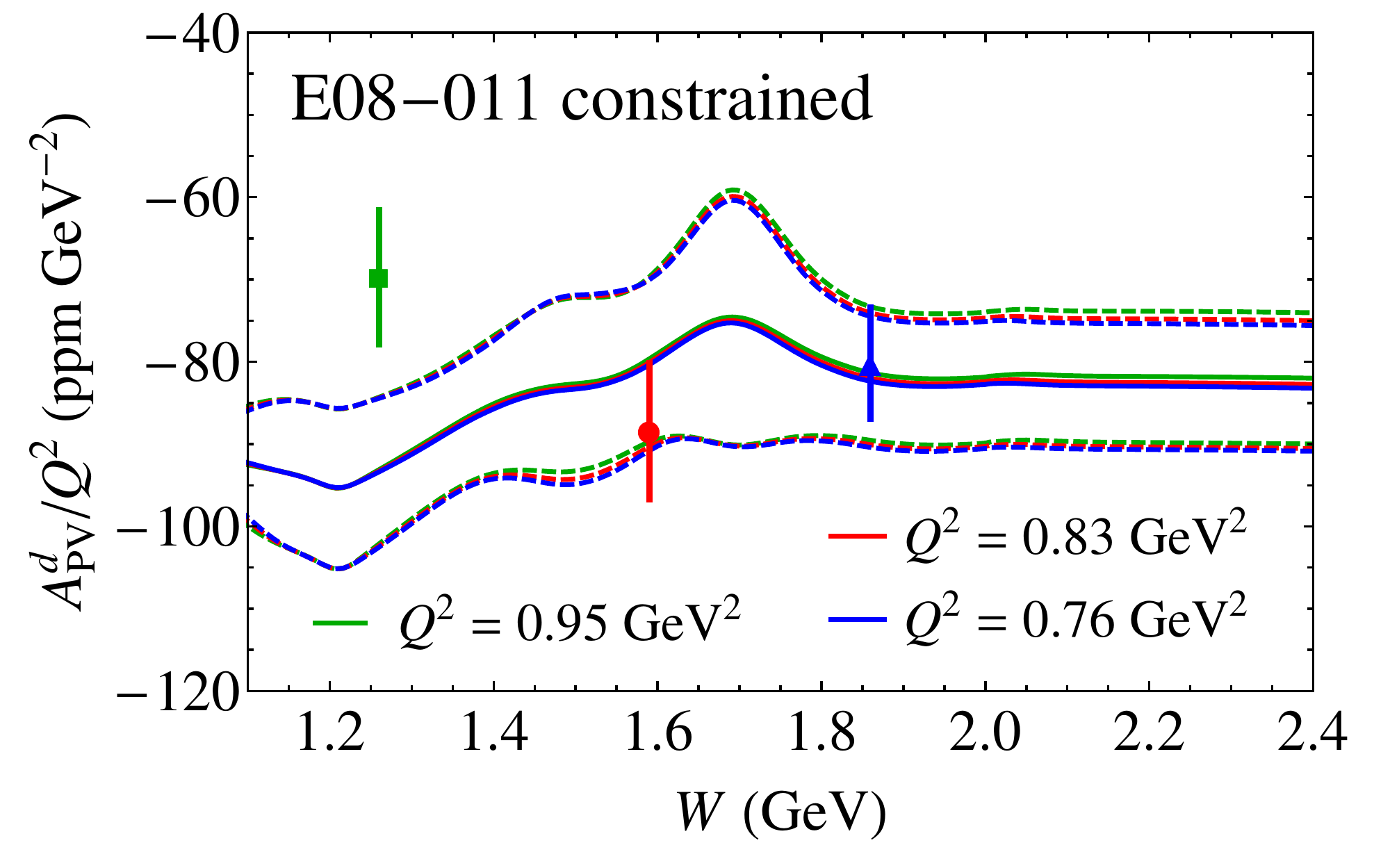}%
\includegraphics[width=3.2in]{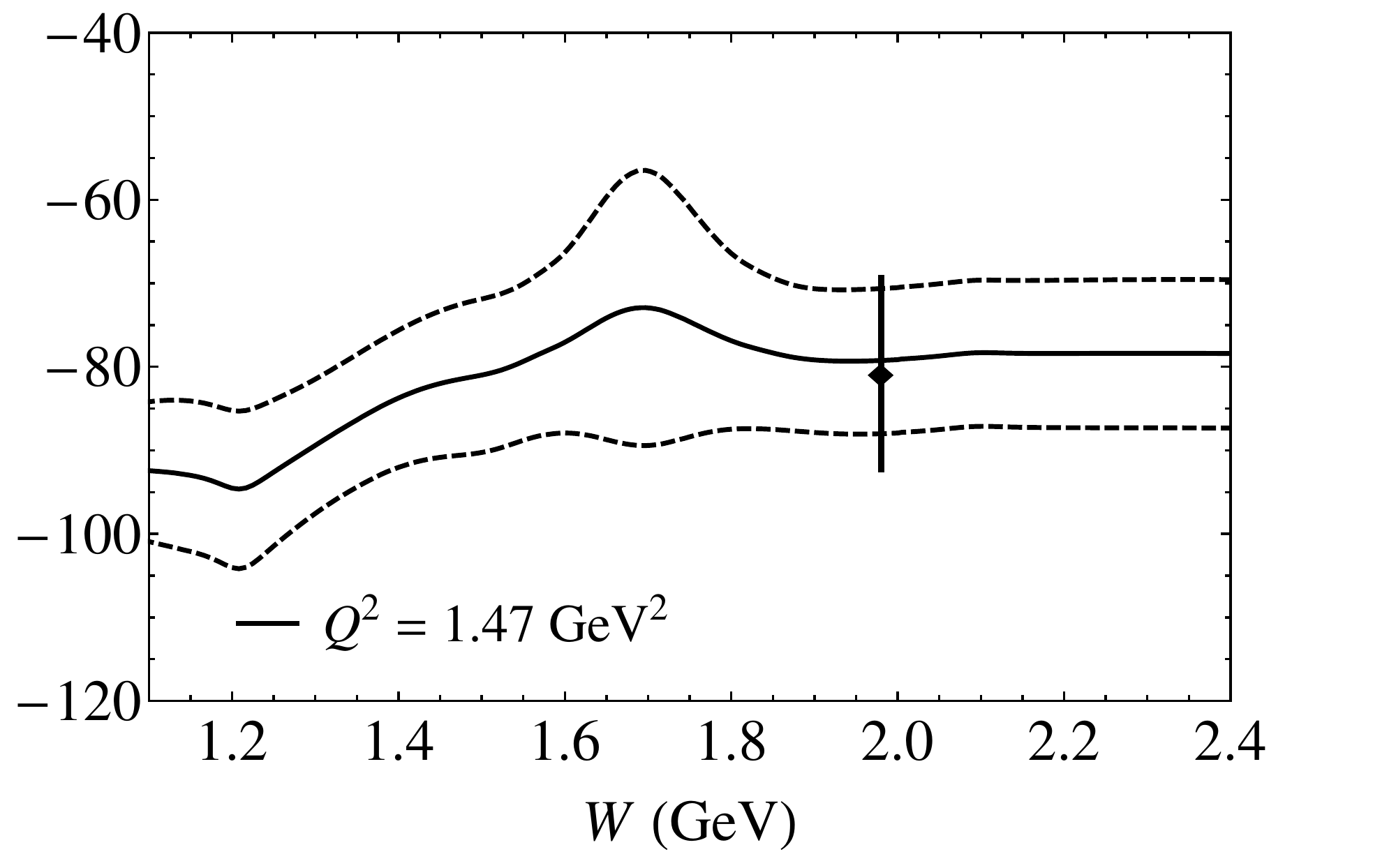}
\caption{(color online)
	As in Fig.~\ref{fig:apvcp}, but with the AJM model
	asymmetries (solid) and their uncertainties (dashed)
	constrained by the E08-011 data \cite{E08011-RES}.
	Note the different scale on the ordinate to that
	in Fig.~\ref{fig:apvcp}.\\}
\label{fig:apvCh}
\end{figure}

The measured parity-violating asymmetry $A_{\rm PV}^d$, scaled by
$1/Q^2$, is shown in Fig.~\ref{fig:apvcp} at $W = 1.26$, 1.59,
1.86 and 1.98~GeV, with $Q^2$ values ranging from 0.76
to 1.47~GeV$^2$.  (The $1/Q^2$ scaling factor enables the various
points to be shown on the same graph.)
The deuteron asymmetries in the AJM model are computed with the
continuum parameters constrained by the DIS region structure
functions computed from global PDFs \cite{Alekhin:2012ig},
as for the proton asymmetry in the previous section
(see Fig.~\ref{fig:apvProMII}).
The resulting fit gives for the transverse continuum parameter
$\kappa_C^T(d) = 0.79 \pm 0.05$, and is in excellent agreement
with the E08-011 data \cite{E08011-RES} for all kinematics, except
at the $\Delta$ region point at $Q^2 = 0.95$~GeV$^2$, where it
lies slightly below the data.
Since the calculation of the $\Delta$ resonance contribution to
$A_{\rm PV}^d$ relies only on isospin symmetry and the conservation
of the vector current, its uncertainty is smaller than that for
higher-mass resonances.  The discrepancy may reflect stronger isospin
dependence of the nonresonant background for $\Delta$ production
\cite{Mukhopadhyay:1998mn}, although the difference is at the
$\lesssim 2 \sigma$ level.  Also, as seen in Fig.~\ref{fig:G0apvProMII}
above, the models agree well with the G0 data \cite{Androic:2012doa}
in the $\Delta$ region, albeit within larger errors.

By using the longitudinal structure function from the global QCD
fit in Ref.~\cite{Alekhin:2012ig}, we find for the longitudinal
continuum parameter $\kappa_C^L(d) = 0.2 \pm 3.4$.  Although the
specific implementation of the CB parametrization \cite{Christy:2007ve}
prevents this uncertainty from being propagated directly into
$A_{\rm PV}^d$, we nevertheless can use the $\kappa_C^{T,L}$
values for the proton to ensure that the uncertainty in the
longitudinal piece is taken into account.
For comparison, we also show in Fig.~\ref{fig:apvcp} the uncertainty
that would be obtained with a similar 100\% error on the continuum
parameters as was assumed by GHRM for the proton, with the VMD+Regge
model \cite{Alwall:2004wk} used for the entire kinematic region
\cite{Gorchtein:2011mz}.
In this case the uncertainties on $A_{\rm PV}^d$ in the
$W \gtrsim 1.8$~GeV region are $\approx 6$ times larger than
the AJM model asymmetries.
Using a reduced 25\% uncertainty on $\kappa_C^T(d)$ results in
asymmetries with a significantly smaller error band, which is
nevertheless slightly larger than in the AJM model.

As a check, the parameter $\kappa_C^{T}(d)$ was also constrained
by performing a $\chi^2$ fit to the E08-011 data points.  This fit
constrains the dominant, transverse continuum parameter to be
$\kappa_C^T(d) = 0.69 \pm 0.13$.
[Omitting the $\Delta$ datum from the fit would yield a marginally
larger value, $\kappa_C^T(d) = 0.72 \pm 0.13$.]
For the longitudinal contribution, the CB parametrization of the
deuteron structure function provides only $F_1^{\gg}$, while
$F_L^{\gg}$ is obtained through the longitudinal to transverse cross
section ratio $\sigma_L^{\gg}/\sigma_T^{\gg}$ [see Eq.~(\ref{eq:LTdef})],
with the deuteron ratio assumed to be the same as for the proton.
Within this parametrization, a direct constraint on $\kappa_C^L(d)$
as for the proton case is therefore not possible.
However, as for the PDF-constrained asymmetry, we can still propagate
the uncertainty on $\sigma_L/\sigma_T$ through the final asymmetry by
including the uncertainties in the $\kappa_C^{T,L}$ values of the
proton which serve as inputs into the $\sigma_L^{\gZ}/\sigma_T^{\gZ}$
ratio.

%
\begin{table}
\begin{center} 
\caption{Parity-violating deuteron asymmetries in the AJM model
	at the kinematics of the E08-011 experiment \cite{Xiaochao:priv, E08011-RES}.
	The asymmetries are computed with the continuum parameters
	$\kappa_C^{T,L}(d)$ constrained by the E08-011 data, or
	by matching to the DIS region described in terms of PDFs.
	Note that the points marked with asterisks ($^*$) are
	predictions.\\}
\begin{tabular}{ c c c | c | c }			\hline
$E$   &  $W$  & $Q^2$	& 
	\multicolumn{2}{c}{$A_{\rm PV}/Q^2$ (ppm GeV$^{-2}$)}
							\\
(GeV) & (GeV) &(GeV$^2$)\ \ & \ \ PDF constraint\ \ 
			& \ \ E08-011 constraint\ \		\\ \hline 
4.9   & 1.26  &  0.95 	& $-93.7_{-9.0}^{+8.8}$
			& $-93.1_{-9.0}^{+8.8}$		\\
4.9   & 1.59  &  0.83 	& $-82.7_{-9.9}^{+9.7}$
			& $-80.1_{-10.3}^{+10.1}$		\\
4.9   & 1.86  &  0.76 	& $-86.2_{-6.9}^{+6.7}$
			& $-82.4_{-8.0}^{+7.9}$		\\
6.1    & 1.98  &  1.47 	& $-84.7_{-6.4}^{+6.2}$
			& $-79.2_{-8.8}^{+8.6}$		\\ \hline
6.1    & 2.03  &  1.28 	& $-84.9_{-6.4}^{+6.2}${}$^{\ (*)}$
			& $-79.7_{-8.6}^{+8.4}${}$^{\ (*)}$	\\
6.1    & 2.07  &  1.09 	& $-85.2_{-6.4}^{+6.2}${}$^{\ (*)}$
			& $-80.3_{-8.3}^{+8.2}${}$^{\ (*)}$	\\
6.1    & 2.33  &  1.90 	& $-82.7_{-6.5}^{+6.3}${}$^{\ (*)}$
			& $-76.5_{-9.3}^{+9.3}${}$^{\ (*)}$	\\ \hline
\end{tabular}
\label{tab:asym}
\end{center}
\end{table}

The resulting asymmetries are again in very good agreement with
the E08-011 data, as is seen in Fig.~\ref{fig:apvCh}.  Moreover,
the uncertainties (dashed curves) are three to four times smaller in the
$W \gtrsim 1.8$~GeV region than those obtained by assuming a 100\%
uncertainty on the parameters, and remain smaller than even for
the reduced, 25\% uncertainty case.
The consistency between the data and the results given by the
constrained expressions gives us confidence in the reliability of
the $\gZ$ structure functions in the AJM model in the region of
low to intermediate $W$ and $Q^2$ that is of greatest importance
for the \regzv calculation.

Finally, the values of the calculated asymmetries and their
uncertainties, using both the resonance region data and the PDF
constraints, are summarized in Table~\ref{tab:asym} at each of
the kinematic points from the E08-011 experiment \cite{E08011-RES}.
In addition, we list the AJM model predictions for $A_{\rm PV}^d$ at
the measured DIS region points at $W > 2$~GeV (marked by asterisks),
which will be discussed further in the next section.

\section{Results}
\label{sec:res}

\subsection{$\gZ$ box corrections for \qwe}

\begin{figure}[t]
\includegraphics[width=5in]{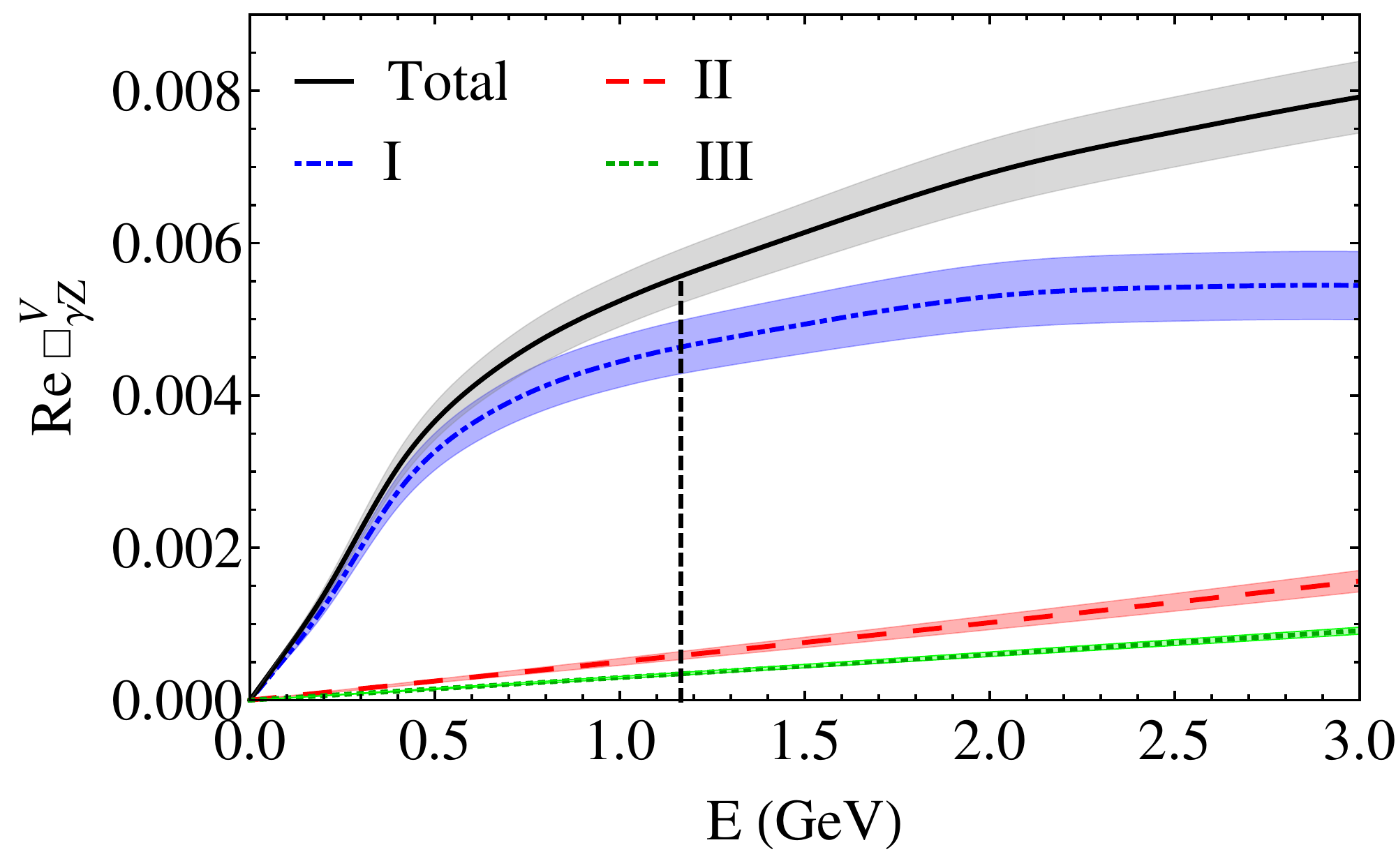}
\includegraphics[width=5in]{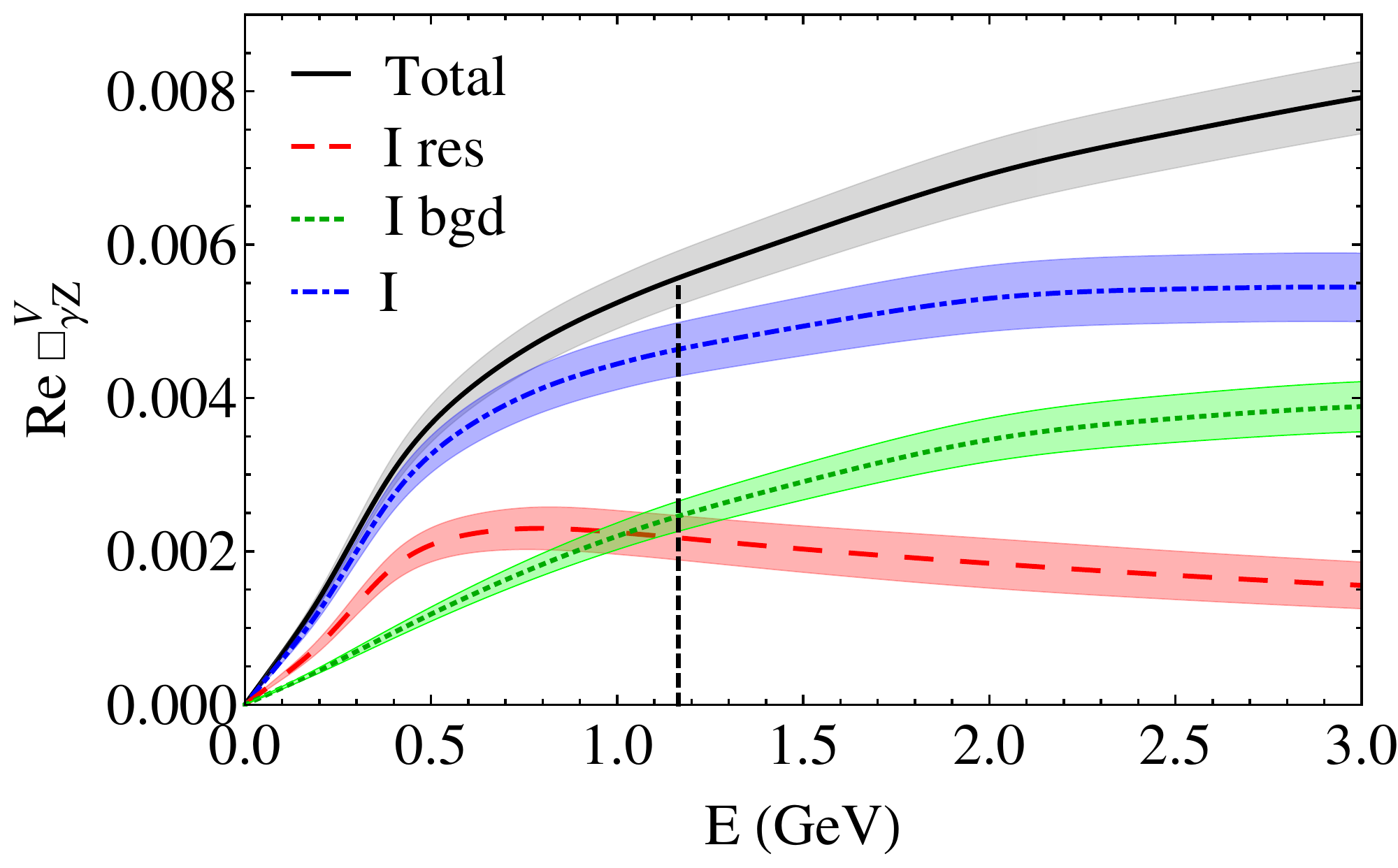}
\caption{(color online)
	Energy dependence of the contributions to \regzv
	from the various regions in $W$ and $Q^2$ displayed
	in Fig.~\ref{fig:W2Q2} in the AJM model (top),
	and the breakdown of Region~I into its resonant
	and nonresonant background components (bottom).}
\label{fig:OurReBox}
\end{figure}

The detailed examination of the $\gZ$ interference structure functions
and their uncertainties, constrained by data in the DIS region and
parity-violating asymmetries in the resonance region, allows us to
compute the \imgzv correction in Eq.~(\ref{eq:ImBoxV}), and through
the dispersion relation (\ref{eq:DR}) to make a reliable determination
of the $\gZ$ box correction to $Q_W^p$.  The dependence of \regzv on
the incident energy $E$ is illustrated in Fig.~\ref{fig:OurReBox},
which also shows the individual contributions of the various $W$ and
$Q^2$ regions in Fig.~\ref{fig:W2Q2}.

At low energy ($E \lesssim 1$~GeV), the total correction \regzv
is dominated by the low-$W$, low-$Q^2$ region (Region~I in
Fig.~\ref{fig:W2Q2}).  As found in earlier analyses
\cite{Gorchtein:2008px, Tjon:2009hf, Sibirtsev:2010zg, Rislow:2010vi,
Gorchtein:2011mz}, the resonant contribution [mainly from the
$\Delta(1232)$ resonance] peaks at around $E \approx 0.7$~GeV,
and gradually decreases at higher energies.  The nonresonant
and resonant components of Region~I are approximately equal at
$E \sim 1$~GeV, with the nonresonant part growing with increasing
energy.  The higher-$W$, higher-$Q^2$ regions play a relatively
minor role in the \bgZv correction, with Regions~II and III
contributing $\approx 20\%$ and 10\% of the total, at $E=3$~GeV,
respectively.

\begin{table}[t]
\caption{Contributions to \regzv from various regions in $W$ and $Q^2$
	in the AJM model (see Fig.~\ref{fig:W2Q2}) at the \qwe
	energy $E=1.165$~GeV.\\}
\begin{tabular}{ c | c }			\hline
Region	  &\ \ \regzv\ ($\times 10^{-3}$)	\\ \hline
I (res)   &  $2.18 \pm 0.29$   			\\
I (bgd)   &  $2.46 \pm 0.20$   			\\
I (total) &  $4.64 \pm 0.35$   			\\
II	  &  $0.59 \pm 0.05$   			\\
III	  &  $0.35 \pm 0.02$   			\\ \hline
Total	  &  $5.57 \pm 0.36$			\\ \hline
\end{tabular}
\label{tab:ReBox}
\end{table}

At the \qwe energy, $E = 1.165$~GeV, the breakdown of the \regzv
correction into its individual contributions is summarized in
Table~\ref{tab:ReBox}.  Including uncertainties from all regions,
the total correction is found to be
\be
\Re e\, \square_{\gZ}^V
	= (5.57 \pm 0.21_{\, \rm [bgd]}
		\pm 0.29_{\, \rm [res]}
		\pm 0.02_{\, \rm [DIS]}) 
\times 10^{-3},
\label{eq:ReBox_final}
\ee
where the uncertainties listed are from the nonresonant background,
the resonances, and the DIS region, respectively.
Adding the errors in quadrature gives
	$\Re e\, \square_{\gZ}^V = (5.57 \pm 0.36) \times 10^{-3}$
at the \qwe energy.  The $\approx 7\%$ relative uncertainty on this
correction remains largely energy independent, even at large energies,
where the contributions from larger $W$ and $Q^2$ become more important;
since the structure functions are constrained by DIS data, the
uncertainty in \regzv does not grow with $E$.

The AJM model value of the $\gZ$ box correction is similar to the result,
	$\Re e\, \square_{\gamma Z}^V = (5.40 \pm 0.54) \times 10^{-3}$,
obtained using the $\gZ$ structure functions from Region~II extended
over all kinematics, as in the GHRM Model~II \cite{Gorchtein:2011mz},
but with the $\kappa_C^{T,L}$ parameters constrained by matching to
the DIS region structure functions \cite{Alekhin:2012ig}.
This constraint renders the uncertainty $\sim$ four times smaller than
that in Ref.~\cite{Gorchtein:2011mz}, but still slightly larger than
in the AJM model calculation.

\subsection{Predictions for parity-violating asymmetries}

\begin{figure}[t]
\includegraphics[width=3.2in]{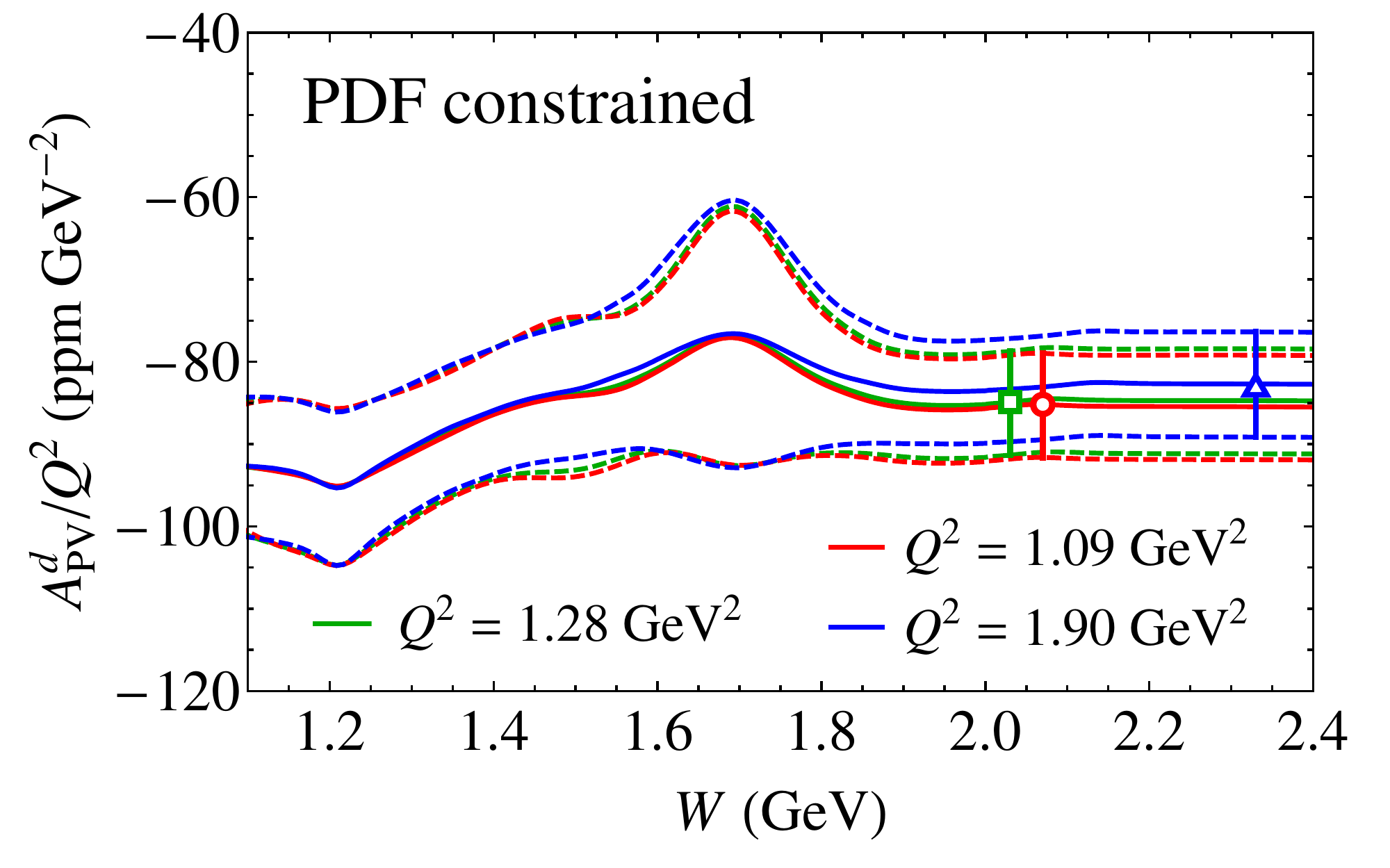}%
\includegraphics[width=3.2in]{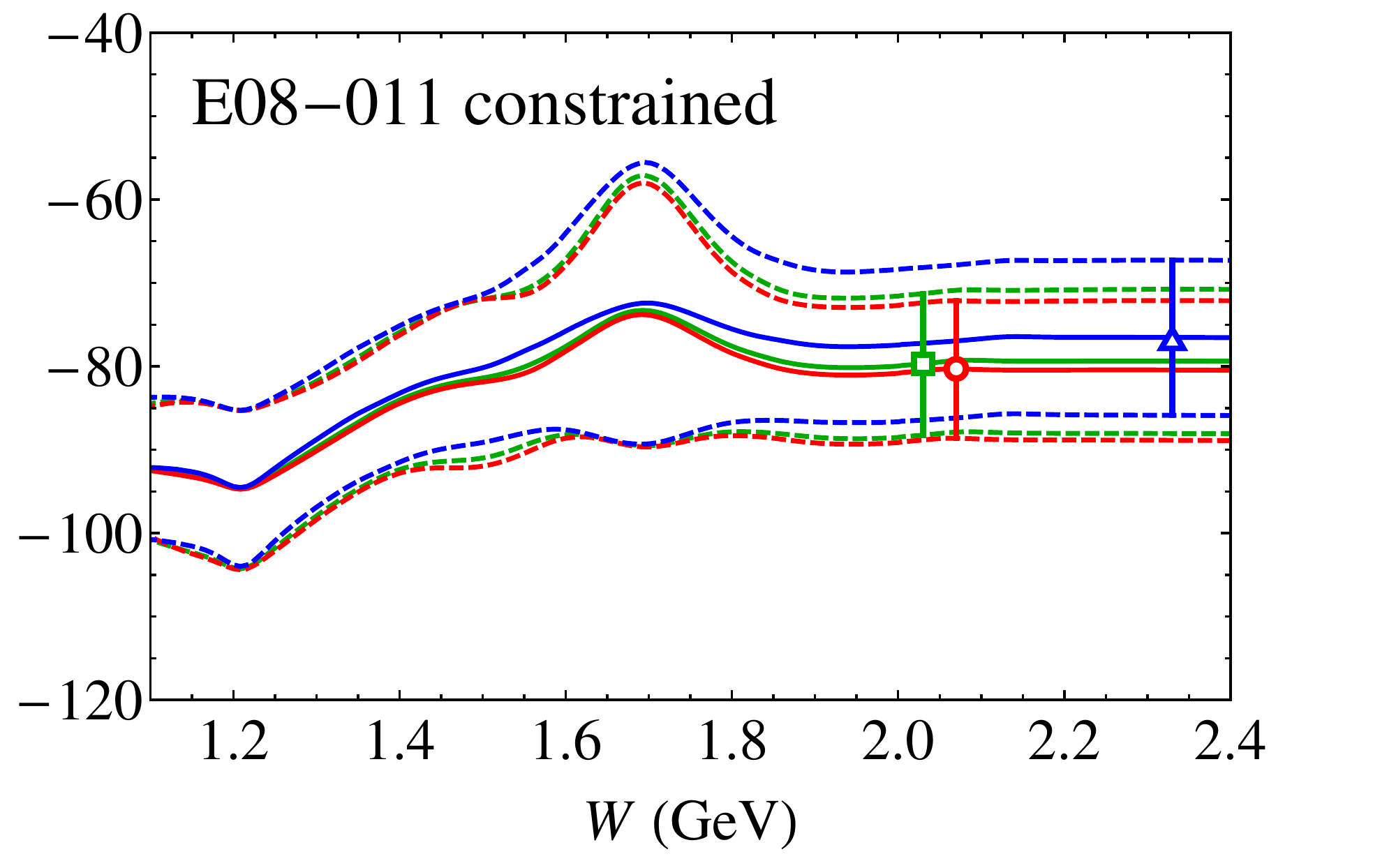}
\caption{(color online)
	Predictions for the parity-violating deuteron asymmetry
	$A_{\rm PV}^d/Q^2$ as a function of $W$ (solid) for the
	DIS region kinematics of the Jefferson Lab E08-011 experiment
	\cite{Xiaochao:priv} at $Q^2 = 1.28$~GeV$^2$ (green),
	1.09~GeV$^2$ (red) and 1.90~GeV$^2$ (blue)
	(see also Table~\ref{tab:asym}).
	The uncertainties (dashed) are computed in the AJM model
	with the continuum parameters $\kappa_C^{T,L}$ constrained
	by DIS structure functions (left), and by the E08-011
	resonance region data (right).
	The predictions at the experimental $W$ values \cite{Xiaochao:priv}
	are shown as pseudo-data points (open symbols).}
\label{fig:Apred}
\end{figure}

The $\gZ$ structure functions can be further constrained by additional
parity-violating asymmetry data from the E08-011 experiment at
Jefferson Lab \cite{Xiaochao:priv, E08011-RES}.  The deep-inelastic region data are
currently being analyzed \cite{Xiaochao:priv}, and the predictions from
the AJM model are shown in Fig.~\ref{fig:Apred} as a function of $W$ for
the three experimental $Q^2$ values (see also Table~\ref{tab:asym}).
The uncertainties on the predictions are computed both by fitting the
continuum parameters $\kappa_C^{T,L}$ to the DIS structure functions
\cite{Alekhin:2012ig} and the E08-011 resonance region data
\cite{E08011-RES}.
The asymmetries with the E08-011 data constraints are marginally
higher than those with the parameters constrained by PDFs, with
slightly larger uncertainties.
As for the resonance region comparison in Figs.~\ref{fig:apvcp} and
\ref{fig:apvCh}, these uncertainties are $\approx$ four to five times smaller
than they would be without the constraints on $\kappa_C^{T,L}$,
assuming 100\% errors along the lines of the proton calculation
in Ref.~\cite{Gorchtein:2011mz}.
The upcoming data will therefore be extremely useful in determining
the uncertainties on the $\gZ$ structure functions and on the
resulting \regzv correction.

\begin{figure}[ht]
\includegraphics[width=4in]{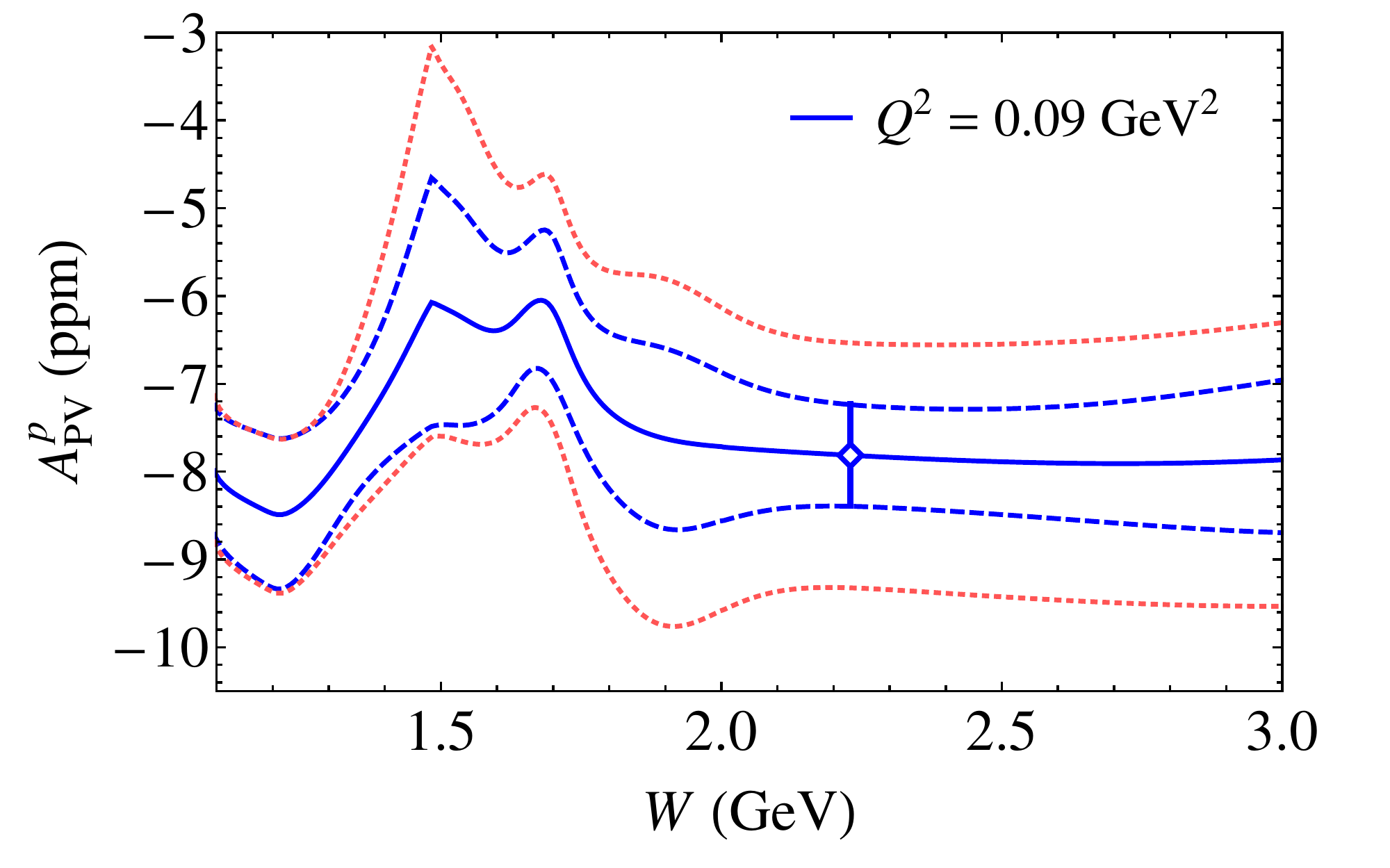}
\caption{(color online)
	AJM model prediction for the proton parity-violating asymmetry
	$A_{\rm PV}^p$ as a function of $W$ for the \qwe inelastic
	measurement \cite{Qweak-inel} at $Q^2 = 0.09$~GeV$^2$
	(solid line and open symbol).
	The AJM model uncertainties (dashed) are compared with those
	from the GHRM model with 100\% uncertainty on the continuum
	parameters (dotted).\\}
\label{fig:Aqwk}
\end{figure}

A further constraint will be provided by the inelastic \qwe measurement
\cite{Qweak-inel}, which was a special run of the \qwe experiment
tuned to the inelastic region at an average $W = 2.23$~GeV.
The AJM model prediction for the proton asymmetry $A_{\rm PV}^p$
and its uncertainty are shown in Fig.~\ref{fig:Aqwk}, where we find
	$A_{\rm PV}^p = (-7.8 \pm 0.6)$~ppm
at the experimental $Q^2 = 0.09$~GeV$^2$ value.
The uncertainty in the AJM model, with the continuum parameters
$\kappa_C^{T,L}$ constrained by the DIS structure functions, is
$\approx$ two times smaller at the inelastic \qwe kinematic point
than that from the GHRM model \cite{Gorchtein:2011mz} without these
constraints.
Note also that in the resonance region, $W \sim 1.5$~GeV, the
uncertainty in the GHRM model almost doubles by taking extrema
values instead of the more conventional addition in quadrature.
The inelastic $Q_{\text{weak}}$, and similar measurements of the
parity-violating inelastic asymmetries, will be valuable for
constraining the $\gZ$ structure functions and the \regzv
corrections in the future.

\section{Conclusion}
\label{sec:conc}

We have performed a comprehensive analysis of the $\gamma Z$ box
contribution to the forward electron-proton elastic parity-violating
asymmetry.  Our primary result is a new determination of the
uncertainty on \regzv{} at the beam energy of the \qwe experiment.
In comparison with previous estimates, we report a significant
reduction in this uncertainty, driven largely by data on structure
functions in the DIS region, and measurements of parity-violating
asymmetries in the resonance region.

To isolate the dependence on the various inputs required in the
evaluation of \regzv, we have divided the dispersion integral into
three kinematic regions.  Region~I, which includes resonance
contributions at low $W$ and $Q^2$, is identified to totally dominate
the value of \regzv.  The total uncertainty is therefore largely
driven by how well the $\gZ$ interference structure functions
$F_i^{\gZ}$ can be constrained in this region.

The resonance region $\gZ$ structure functions are determined by
an isospin transformation of the corresponding $\gg$ structure
functions.  The input $F_i^{\gg}$ functions are determined by a fit
\cite{Christy:2007ve} to the world's inclusive electron-nucleon
scattering data in terms of resonance contributions and a nonresonant
background.
For the resonance components, the isospin transformation can be
performed using the conservation of the vector current and the
isospin dependence of the couplings, as reported by the PDG,
with relatively modest contribution to the overall uncertainty.
For the background, following the approach of
Ref.~\cite{Gorchtein:2011mz}, the transformation is estimated
using a prescription based on the VMD model \cite{Alwall:2004wk}.
For the low-mass vector meson components the isospin rotation is
determined by isospin symmetry of the electroweak interactions,
while the transformation of the high-mass continuum part is not
fixed within the VMD formalism, and consequently contributes a
larger uncertainty.

At larger $Q^2$ values ($Q^2 > 1.5$~GeV$^2$) the continuum piece
totally dominates the nonresonant background.  We use this fact to
constrain the continuum component of the isospin rotation by matching
this to the DIS structure functions in the transition region.
The model dependence from using a particular continuum form at lower
$Q^2$ (away from the PDF constraint) is less important, since this
region is dominated by the low-mass vector mesons $\rho$, $\omega$
and $\phi$.  It is the constraint on this rotation that drives the
significant reduction in uncertainty in the present AJM model as
compared to that reported by GHRM \cite{Gorchtein:2011mz}.

Combined with the relatively well-determined contributions from
Regions~II and III at higher $W$ and $Q^2$ (see Fig.~\ref{fig:W2Q2}),
we find the final value for $\gZ$ correction to be
	\regzv$ = (5.57 \pm 0.36) \times 10^{-3}$.
Importantly, this precision maintains confidence in the
interpretation of the \qwe experiment as a standard model test.

The reliability of our constraint procedure has been confirmed by
a comparison with the corresponding inclusive $\gZ$ interference
asymmetries recently measured on the deuteron by the E08-011
experiment at Jefferson Lab \cite{E08011-RES}.
Conversely, using the E08-011 resonance region data as a constraint
on the $\gZ$ structure functions, the resulting asymmetries are found
to be very similar to those in the AJM model with the PDF constraints,
albeit with slightly larger uncertainties.  Upcoming data on the
deuteron asymmetry in the DIS region \cite{Xiaochao:priv} should
reduce these uncertainties.

Beyond this, the most promising means by which one could further
constrain the $\gZ$ structure functions would be to perform a
systematic experimental study of parity-violating electron scattering
on hydrogen across Region~I.  While the recent deuterium measurements
\cite{E08011-RES} have proven useful in providing confidence in the
procedure of matching to PDFs at intermediate $Q^2$ and $W$, because
the deuteron requires a knowledge of the neutron structure function
as well as of the proton, this has limited value as a means to reduce
the uncertainty in $F_i^{\gZ}$.  A dedicated study of the proton itself
would directly constrain the model and lead to a reduction in the
uncertainty of the radiative correction arising from the $\gZ$ box.

\section*{Acknowledgements}

We thank S.~Alekhin, R.~Carlini, C.~Carlson, M.~Dalton, M.~Gorshteyn,
K.~Meyers, R.~Michaels and X.~Zheng for helpful discussions and
communications.
N.~H. and P.~B. thank the Jefferson Lab Theory Center for
support during visits where some of this work was performed.
P.~B. and W.~M. thank the CSSM/CoEPP for support during visits to
the University of Adelaide.
This work was supported by NSERC (Canada), DOE Contract No. DE-AC05-06OR23177,
under which Jefferson Science Associates, LLC operates Jefferson Lab;
DOE Contract No. DE-FG02-03ER41260, and the Australian Research Council
through an Australian Laureate Fellowship.

%
%

\end{document}